\documentclass[prb,nobibnotes,altaffilletter,amsmath,amssymb,amsfonts]{revtex4}

\usepackage{inputenc}
\inputencoding{ansinew}
\usepackage[T1]{fontenc}
\usepackage{graphicx}
\usepackage{palatino} 
\usepackage{mathpazo} 
\usepackage{bm} 
\usepackage{color}
\newcommand*{\parti}[2]{\frac{\partial#1}{\partial#2}}
\newcommand*{\diff}[2]{\frac{\mathrm{d}#1}{\mathrm{d}#2}}
\renewcommand*{\(}{\left(}
\renewcommand*{\)}{\right)}

\renewcommand*{\]}{\right]}
\newcommand*{\bkww}{\beta}

\begin{document}

\title{The dynamic thermal expansivity of liquids near the glass transition.}
\author{Kristine Niss}
\author{Ditte Gundermann}
\author{Tage Christensen}
\author{Jeppe C. Dyre}
\address{DNRF Centre Glass and Time, IMFUFA, Department of
  Sciences, Roskilde University, Postbox 260, DK-4000 Roskilde,
  Denmark
 } \date{\today}

\begin{abstract}
  Based on previous works on polymers by Bauer {\it et al.} [Phys,
  Rev. E (2000)], this paper describes a capacitative method for
  measuring the dynamical expansion coefficient of a viscous
  liquid. Data are presented for the glass-forming liquid tetramethyl
  tetraphenyl trisiloxane (DC704) in the ultraviscous regime.
  Compared to the method of Bauer {\it et al.} the dynamical range has
  been extended by making time-domain experiments and by making very
  small and fast temperature steps. The modelling of the experiment
  presented in this paper includes the situation where the capacitor
  is not full because the liquid contracts when cooling from room
  temperature down to around the glass-transition temperature, which
  is relevant when measuring on a molecular liquid rather than 
  polymer.
 \end{abstract}

\maketitle
\begin{figure}
  \centering
  \includegraphics[width=0.8\linewidth]{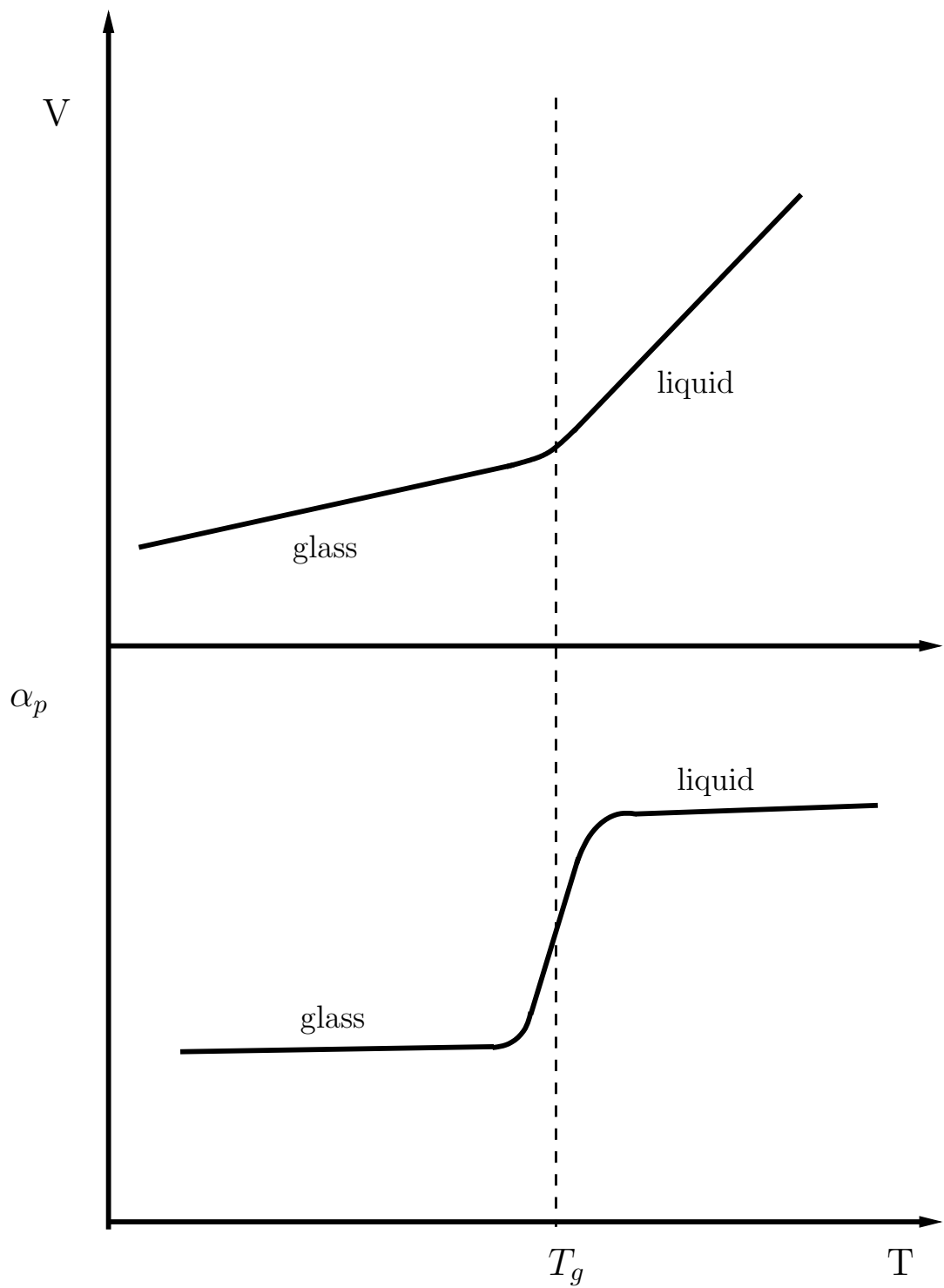}
  \caption{Illustration of the temperature dependence of volume and
    expansion coefficient of a liquid in the vicinity of the
    glass-transition. Upon cooling the expansivity decreases  abruptly at the glass
    transition. This gives rise to a kink in the temperature
    dependence of the volume. These features are the original
    signatures of the glass transition.    }
  \label{figur1figur}
\end{figure}

{}
The glass transition occurs when the configurational degrees of
freedom of a liquid are frozen in. Below the glass transition
temperature, $T_g$, only isostructural contraction takes place as
temperature is decreased further. The measured thermal expansion
coefficient, $\alpha$ (and heat capacity, $c_p$) are therefore lower
in the glass than in the equilibrium liquid. This change of the
thermal expansion coefficient (and the heat capacity), is probably the
most classical signature of the glass transition, and a figure
illustrating this change (see Fig.  \ref{figur1figur}) is almost
inevitably the starting point of introductory talks or texts on the
glass transition (eg. Ref. \onlinecite{angell00,dyre06}).

The change in the heat capacity at the glass transition $\Delta c_p =
c_{p,liq} -c_{p,glass}$ has been studied extensively and is widely
believed to play a role for the dynamics of liquids close to the glass
transition. The change in expansion coefficient $\Delta \alpha_{p} =
\alpha_{p,liq} -\alpha_{p.glass}$ has received less attention,
but is of similar importance. This is seen, for
instance, in the literature related to the Prigogine-Defay ratio, a
dimensionless number characterizing the glass
transition \cite{davies53,prigogine54,goldstein63,moynihan78,ellegaard07}.

The glass is an out-of-equilibrium state and therefore the values of
the thermodynamic derivatives are not rigorously well defined. They
depend on cooling rate and also on the time spent in the glassy
state. Contrary to this the linear response of the metastable
equilibrium liquid state is well defined and history independent
\cite{ellegaard07}. The linear expansion coefficient of a viscous
liquid close to its glass transition is dynamic, that is time (or
frequency) dependent with short times giving a low (glass-like) value,
$\alpha_{p,fast}$, while long times give a higher liquid value,
$\alpha_{p,slow}$. The difference between these two levels $\Delta
\alpha_{p,lin} = \alpha_{p,slow} -\alpha_{p,fast}$ thus gives
well-defined information on the configurational part of the expansion
coefficient. Likewise $\Delta c_{p,lin} = c_{p,slow} -c_{p,fast}$ is
well defined.

The relaxation between the fast and the slow response takes place on a
certain time scale which is temperature dependent. Considered in this
way the measurement of the expansion coefficient can be viewed as a
type of spectroscopy, which gives both a relaxation time and a
spectral shape analogous to other methods like dielectric spectroscopy
or mechanical spectroscopy. The study of the temperature dependence of
relaxation times and of the spectral shape of different response
functions is vital for understanding the viscous slowing down. There
is a general belief that the liquid has a relaxation time, which is
fairly well defined independent of probe, but also suggestions that
different processes may decouple from each other at low
temperatures \cite{Angell1991}. 

There are good scientific reasons to study the dynamic linear
expansion coefficient,  but almost no data of this type are to be
found in literature.  The time-dependent expansion coefficient can be
found by studying the change in volume as a function of time after a
temperature step. Such volume
relaxation experiments are very classic in glass science and still
important \cite{kovacs58,greiner89,kolla05,svoboda06}. However, 
volume relaxation experiments are traditionally performed as
non-linear aging experiments, \emph{i.e.}, with large amplitudes in the
temperature jump. This type of experiment gives information on the
relaxation of the configurational degrees of freedom, but the
expansion coefficient and its characteristic time scale cannot be
determined, because the results depend on the amplitude and sign of
the temperature jump. For sufficiently small temperature steps this is
not the case, defining the linear response regime. 

The only linear dynamic data we are aware of were reported about a decade ago
by Bauer \emph{et al.}  \cite{bauer00,bauer01} followed by a paper by
Fukao and Miyamoto \cite{fukao01}.  These papers reported frequency-domain
measurements on thin polymer films, performed with temperature
scans at a couple of fixed frequencies, covering 1.5 decade of
the dynamics. The measurements were pioneering, but 1.5 decade is not
very much for studying relaxation in
viscous liquids, because the relaxation is extremely temperature
dependent and quite ``stretched'', which means that 
even at one fixed temperature the relaxation covers several decades.

The technique developed by Bauer \emph{et al.} is based on a principle
where the sample is placed in a parallel plate capacitor such that it
is the sample that maintains the spacing between the plates. Thus a
change in sample volume in response to temperature change leads to a
change of the capacitance. This principle is also used in the present
work. The advantage of this technique is that capacitance can be
measured with high accuracy and it is this accuracy which makes linear
experiments possible.

The use of sample-filled capacitors for measuring an expansion
coefficient is not unique and it has been done by others before and
after Bauer \emph{et al.} (see
eg. Ref. \onlinecite{meingast96,fukao99,serghei06,oh09}) in
capacitative scanning dilatometry, \emph{i.e.} working in a
temperature ramping mode. Capacitative scanning dilatometry has to our
knowledge never been used on simple liquids. It is particularly useful
for studying thin polymer films because the signal gets better with a
thin sample.  The technique has been used for determining the
glass-transition temperature for example as a function of film
thickness \cite{fukao99,serghei06} or as a function of cooling rate
\cite{meingast96}. The main focus of these papers is on the
temperature dependence of the expansion coefficient, while little
attention is given to the absolute values. There have been no studies
of the dynamics since the pioneering work of Bauer and no attempts to
extend the dynamical range.

To the best of our knowledge there are no measurements of the
dynamic linear expansion coefficient of molecular liquids. The reported
data from scanning dilatometry and non-linear volume relaxation
are also mainly for polymers, while data on molecular liquids
is relatively scarce. This may be due to the higher technological
importance of polymers. It is probably also 
related to the fact that working with molecular liquids requires other
experimental conditions, meaning that techniques developed for
polymers are not always directly applicable to liquids. 

This paper gives a description of an experimental method developed for
measuring the dynamical expansion coefficient of a viscous. As
mentioned, the principle is based on the capacitive technique by Bauer
{\it et al.}  \cite{bauer00,bauer01}. The method is modified in three
respects compared to the work of Bauer {\it et al.}: 1) The modelling
takes into account the situation where the capacitor is not full,
which is relevant when measuring on a molecular liquid rather than on
a polymer. 2) The experiment is performed in the time domain using a
very fast temperature regulation, which gives a dynamical range of
more than four decades. 3) The sensitivity is enhanced by using a
capacitance bridge with a very high resolution. This makes it possible
to measure the response following very small temperature steps,
ensuring that the response is close to perfectly linear. As an
application of the technique the paper presents data on the
glass-forming liquid tetramethyl tetraphenyl trisiloxane (DC704) in
the ultraviscous regime.

\section{Response functions with consistent dimensions}\label{sec:response}
In a linear response experiment, the response of a system to an
external perturbation is studied. If the perturbation is small the
output is assumed to be linearly dependent on the input. The formalism
to describe this is well known. However, different formulations can be
used, and the version used in this work when converting the measured
time-domain response to the frequency-domain response function is
maybe not the most common one. The formalism used here has the
advantage that the time-domain response function and the
frequency-domain response function have the same dimension and there
is no differentiation involved when transforming between the two. This
section gives a summary of the response function formalism used
including a comparison to the standard formalism.

The fundamental assumption is that the output depends linearly on the
input. The most general statement is that the change in input $d
I(t')$ at time $t'$ leads to a contribution in output $d O(t)$ at time
$t$:
\begin{equation}\label{lin_response}
dO(t)=R(t-t')dI(t').
\end{equation}
It is here assumed that the change in output only depends on the time difference $(t-t')$. Causality implies that 
\begin{equation}
R(t)=0 \textrm{ for } t< 0.
\end{equation}
Integrating on both sides of Eq. (\ref{lin_response}):
\begin{eqnarray}\nonumber
O(t)=\int^t _{- \infty} R(t-t')dI(t'),
\end{eqnarray}
and substituting $t''=t-t'$  and writing $\dot{I}(t)=\dfrac{dI(t)}{dt}$
\begin{eqnarray}\nonumber
 O(t)&=&-\int_{\infty}^0 R(t'')\dot{I}(t-t'')dt''.
\end{eqnarray}
Changing $t''$ to $t'$:
\begin{eqnarray}\label{lin_response2}
 O(t)=\int^{\infty} _0 R(t')\dot{I}(t-t') dt'.
\end{eqnarray}

 If the input is a Heaviside function:
 \begin{eqnarray}I(t)=I_0H(t)=I_0\left\{\begin{array}{rl}
           0 & \textrm{for } t\leq 0 \nonumber\\
           1 & \textrm{for } t>0\nonumber
          \end{array}\right. 
\end{eqnarray}
 then
 \begin{equation} \label{output1}
 O(t)=I_0\int^{\infty}_0 R(t')\delta(t-t')dt'=I_0R(t),
 \end{equation}
 and it is seen that $R(t)$ is the output from a Heaviside step input.

Linear response can also be studied in the frequency domain. In the
case of a harmonic oscillating input $I(t)=I_0 e^{i(\omega t+\phi_I)}$,
the output $O(t)=O_0e^{i(\omega t+\phi_O)}$ will be a periodic signal
with the same frequency $\omega$, but there will be a phase shift of
the output relative to the input. From Eq. (\ref{lin_response2})
the output is
\begin{eqnarray}
 O(t)&=&\int^\infty_0 R(t')i\omega I_0e^{i\phi_I}e^{i\omega(t-t')}dt'\nonumber \\
      &=&I_0 e^{i\omega t} e^{i\phi_I}i\omega \int^\infty_0R(t')e^{-i\omega t'}dt'\nonumber \\
      &=&I(t)R(\omega),\nonumber
\end{eqnarray}
where $R(\omega)$ is the frequency domain response function, which is given by the Laplace transform of $R(t)$ times $i\omega$:
\begin{equation}\nonumber
 R(\omega)=i\omega \int^\infty_0R(t')e^{-i\omega t'}dt'.
\end{equation}
The linear response relation is often expressed in an alternative
formulation where the linearity assumption is expressed by
\begin{eqnarray}\nonumber
O(t)&=&\int^t_{-\infty} \mu(t-t')I(t')dt',
\end{eqnarray}
where $\mu$ is sometimes called the memory function, but it is also
sometimes called the response function. The use of the word response
function for $\mu(t)$ is somewhat inconvenient because it has a
different dimension compared to the frequency-domain response function
$R(\omega)$. Substituting again ($t''=t-t'$) and changing $t''$ to
$t'$
\begin{eqnarray}\nonumber
O(t)&=&\int_0^\infty \mu(t')I(t-t')dt'.
\end{eqnarray}
Applying a Heaviside input again
\begin{eqnarray}\label{output2}
O(t)&=&\int^\infty_0\mu(t')I_oH(t-t')dt'\nonumber\\
&=&I_0\int^t_0\mu(t')dt'.
\end{eqnarray}
From Eq. (\ref{output1}) and (\ref{output2}) we have
\begin{equation}\nonumber
 R(t)=\int_0^t\mu(t')dt',
\end{equation}
and therefore
\begin{equation}\label{hej3}
 \dfrac{dR(t)}{dt}=\mu(t).
\end{equation}
In the memory function formalism the frequency domain response is
again found by inserting a harmonic oscillating input. In this case
the result becomes 
\begin{equation}\nonumber
 R(\omega)= \int^\infty_0\mu(t')e^{-i\omega t'}dt'=\int^\infty_0\dfrac{dR(t)}{dt}e^{-i\omega t'}dt',
\end{equation}
where the last equality comes from inserting Eq. (\ref{hej3}).  This
expression is formally equivalent to Eq. (\ref{Rw}) which can be shown
by integration by parts and by invoking $R(t=0)=0$. However, when
converting data in practice Eq. (\ref{Rw}) has the advantages that
differentiation of the time domain data is avoided.  It is always good to avoid differentiation of
numerical data because it introduces increased noise. Moreover, if we
introduce an ``instantaneous'' response in terms of  $R(t\rightarrow
0)\neq 0$ corresponding to very short times where we can not measures the time dependence of
the response, then this information would be lost by differentiation.  

\section{Principle, design and procedure}\label{sec:design}
The method requires that there is a simple relation between sample
density and dielectric constant.  The dielectric constant in general
has two contributions: atomic polarization and rotational polarization
\cite{bottcher}.  The atomic polarization is due to the displacement
of the electron cloud upon application of a field. This contribution
is governed by the microscopic polarizability of the molecule, $x$
(usually called $\alpha$, but $\alpha$ is reserved for the expansivity
in this paper).  The atomic polarizability can be assumed to be
temperature and density independent in the relevant range. This means
that the desired simple relation between density and dielectric
constant can be obtained when the atomic polarization is the only
contribution.

The rotational polarization is due to rotation of the permanent
dipoles in the sample. This contribution is relevant when the liquid
has a permanent dipole moment and mainly at frequencies lower than or
comparable to the inverse relaxation time of the liquid. The
rotational contribution gives the dielectric signal monitored in
standard dielectric spectroscopy. The rotational
polarization is temperature-, density- and frequency-dependent, and it
is therefore non-trivial to relate the density to the dielectric
constant when rotational polarization is present.  Therefore, in capacitative dilatometry
it is a contribution one would like to avoid. It is sometimes
assumed that the high frequency plateau value of the dielectric
constant measured in dielectric spectroscopy contains only atomic
polarization and that it corresponds to the square of refraction index
$n^2$. However, there is also fast (``glass-like'') contribution to
the rotational part of the polarization. The fast rotational
contribution will dominate over the geometric effects even at high
frequencies if the sample has a high dipole moment. This was
demonstrated in Ref. \onlinecite{niss05}. To minimize the rotational
contribution two things are done: 1) Only liquids with very small
dipole moment are studied - \emph{i.e.} liquids in which the atomic
polarization is dominant at all frequencies and temperatures. 2) These
liquids are only studied at frequencies much higher than the inverse
relaxation time. In the data reported in this paper the measuring frequency
is 10 kHz and the relaxation time is 100 seconds or more.

The cell is a capacitor made of circular copper plates of 1~cm
diameter and 1~mm thickness, with a 50~$\mu$m spacing. The separation
is kept by four 0.5~mm~x~0.5~mm and 50~$\mu$m thick Kapton
spacers. The spacing between the capacitor plates is filled with the
sample liquid. The thin spacing results in a reasonably large
dielectric signal (empty capacitance is 14~pF) despite the small
size. The thin spacing moreover makes it possible to heat or cool the
sample fast, even though the heat diffusion in the sample liquid is
slow compared to the heat diffusion in the copper plates.

\begin{figure}
  \centering
  \includegraphics[width=0.8\linewidth]{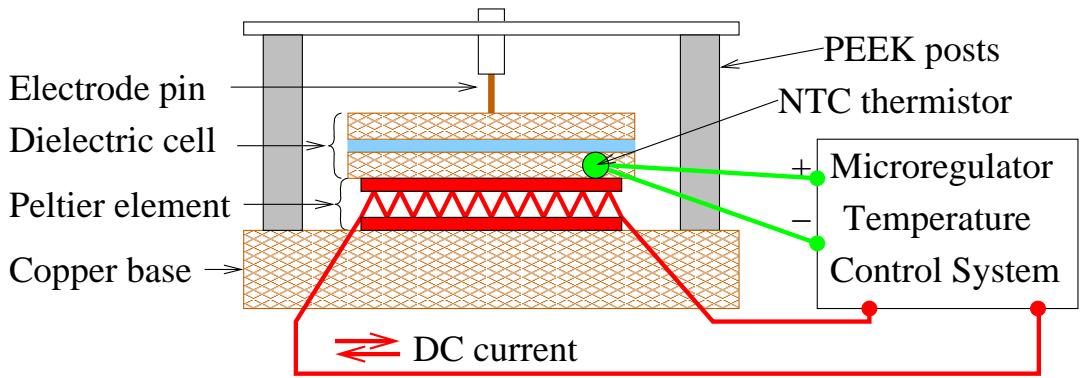}
  \caption{Schematic drawing of the dielectric measuring cell with the microregulator.
The liquid is deposited in the 50~$\mu$m gap between the discs of
the dielectric cell. The Peltier element heats or cools the dielectric cell,
depending on the direction of the electrical current powering the element.
The current is controlled by an analog temperature-control system that receives
temperature feedback information from an NTC thermistor embedded
in one disc of the dielectric cell (reproduced from
    Ref. \onlinecite{hecksher10}).}\label{fig:setup}.
\end{figure}

The cell is integrated with a microregulator, which is a tiny
temperature regulator based on an NTC-thermistor (placed in the lower
cupper plate of the capacitor-cell), a Peltier-element acting as a
local source of heating and cooling, and an analog PID-control. The
integrated cell and microregulator are placed in our main
kryostat. With this setup the temperature of the sample can be changed
by steps of up to 2~K within less than 10~s and the
temperature can be kept stable is within a few micro Kelvin over days and weeks. The
cell is shown in Fig. (\ref{fig:setup}) and the whole system of the main
kryostat and the microregulator is described in detail in
Ref. \onlinecite{igarashi08}.

The principle of the experiment is to make an ``instantaneous'' step
in temperature and subsequently measure the capacitance at a fixed
frequency as a function of time. From the capacitance we calculate the
time-dependent expansion coefficient.  In order for the temperature
step to be ``instantaneous'' compared to the time scale of the
relaxation we need the relaxation time to be 100~s or longer. This
means that the measurements are performed at or below the
conventional glass-transition temperature. Nevertheless, it is
important to emphasize that, the liquid is in
equilibrium when the experiment is performed because we wait at least
five relaxation times whenever stepping to a new temperature before
making a measurement. The measurements themselves also must be
carried out over five relaxation times in order to obtain the
relaxation curve all the way to equilibrium. All together, it takes
days and sometimes even weeks to take a spectrum at a given
temperature. This means that the experiment would be impossible without the stable
temperature control ensured by the microregulator.

The relaxation time of viscous liquids close to the glass transition
is extremely temperature dependent. We therefore need to make small
temperature steps in order for the measured response to be
linear. This means that the change in volume and thereby the measured
capacitance is very small, the relative changes in capacitance
$\textrm{d}C/C$ are of order 
10$^{-4}$. We use an AH2700A
Andeen Hagerling ultra-precision capacitance bridge, which
measures capacitance with an accuracy of 5~ppm and true resolution of
0.5~attoFarad in the frequency range 50~Hz-20~kHz. The capacitance is
measured every second at 10~kHz.

The sample used is liquid at room temperature and the capacitor is
filled by letting the liquid suck in using the
capillary effect. Complete filling is checked by measuring the
capacitance before and after filling, comparing to the measured
dielectric constant measured at the same temperature with a larger
capacitor (which is easy to fill).

\section{Geometry and boundary conditions}\label{flow}

In order to model the relation between the measured change in
capacitance and the expansion coefficient some 
assumptions must be made regarding the behavior of the liquid during the
experiment. In this section we describe these assumption and the
arguments on which they are based.  

The capacitor is filled completely at room temperature with a low
viscosity molecular liquid. The measuring temperatures (close to and
below the conventional glass-transition temperature) are typically around 100
degrees below room temperature for these types of liquids. The cooling
makes the liquid contract in the radial direction because the distance
between the plates is maintained by the spacers (which have a much
smaller expansion coefficient). This has the consequence that the
capacitor is not completely filled at the temperatures where the
measurements take place. This gives rise to a difference
compared to the measurements done on polymers in earlier work
\cite{bauer00,bauer01}, a difference which 
is taken into account when calculating the relation between the
expansion coefficient and the change in capacitance in the following
section.

The liquid contracts/expands radially as long as it has low viscosity,
but the situation changes when the liquid gets ultraviscous. At
high viscosities the liquid gets clamped between the plates due to the
small distance between them. This has the consequence that the liquid
can no longer contract/expand upon cooling/heating by flowing radially, but will
contract/expand vertically and pull/push the plates changing the
distance between them. This effect is the basis for the
measurement, because the vertical expansion makes the capacitance
change, and we calculate the expansion from the change in capacitance.

The distance between the plates is kept by the Kapton spacers at high
temperatures (and long times) when the sample liquid flows. However,
at times where the sample cannot flow, it is the sample, not the
Kapton spacers which determines the distance. This is true because
Kapton has a stiffness \cite{davidson92} of the same order of
magnitude as the sample (in the GPa-range), but only takes up
approximately 1\% of the area between the plates.

The temperature change gives rise to an internal pressure, which is
released by pressure diffusion via viscous flow.  The characteristic
time $\tau_{flow}$ of the radial flow between two plates of fixed
distance $l$ can be estimated by the following argument. A temperature
step of $\Delta T$ initiates an internal pressure $\Delta p = K_T
\alpha_p \Delta T$ in the liquid. This creates a radial flow that
eventually discharges the surplus volume $\Delta V= \Delta T \alpha_p
\pi R^2 l$. Although the volume flows in the radial direction we may
as a crude estimation take the volume velocity, $\dot V$, of planar
Pouiseuille flow\cite{lautrup} $\dot V = \frac{\Delta p}{12 \eta L}W
l^3$, where $L$ (the dimension in the direction of the flow) can be
taken as $R$, and $W$ (the dimension perpendicular to the flow) can be
taken as $2 \pi R$. The characteristic discharge flowtime then becomes
$\tau_flow=\frac{\Delta V}{\dot V}= 6\frac{\eta}{K_T}(\frac{R}{l})^2$.
The bulk modulus and the shear modulus are of the same order of
magnitude.  It follows that the Maxwell relaxation time is roughly
given by $\tau_{M}\simeq \eta/ K_T$ and that $\tau_{flow}\propto
(R/l)^2 \tau_{M}$.  In the experiment we have $l$=50~$\mu$m and
$R=5$~mm from which it follows that the radial flow time is ten
thousand times longer than the Maxwell time. The alpha relaxation time
is roughly given by the Maxwell time, the flow time will be more than
ten days when the alpha relaxation time is one hundred seconds. This
means that the liquid can be considered as radially clamped in the
region we study (where all relaxation times are longer than 100
seconds). The transition between the radial flow and the clamped
situation can be seen in dielectric constant when it is measured as a
function of temperature, and the observed behavior is consistent with
the above estimate.

The expansion coefficient we study with the boundary conditions
described above is not the conventional isobaric expansion
coefficient, $\alpha_p=\left.\frac{1}{V}\parti{V}{T}\right|_{p}$,
because the liquid is clamped in two directions and only free to move
in one direction. We call this expansion coefficient the longitudinal
expansion coefficient, in analogy to the longitudinal modulus,
(another name for it could be the iso-area expansion coefficient). It
is expressed by
$\alpha_l=\left.\frac{1}{V}\parti{V}{T}\right|_{A}=\left.\frac{1}{l}\parti{l}{T}\right|_{A}$,
where $A$ is the constant area and $l$ is the dimension which is free
to respond to the temperature change. The longitudinal expansion
coefficient is related to the isobaric expansion coefficient
$\alpha_p$ via the following relation
\begin{equation}
  \nonumber
  \alpha_l(\omega)=\frac{1}{1+\frac{4G(\omega)}{3K_T(\omega)}}\alpha_p(\omega).
\end{equation}
Where $G$ is the shear modulus and $K_T$ is the isothermal bulk
modulus, which are both dynamic \emph{i.e.} frequency or time
dependent as are the thermal expansion coefficients. 

From this expression we see that $\alpha_l$ is smaller than
$\alpha_p$, except at low frequencies (long times, or high
temperatures) where $G=0$ which implies $\alpha_l=\alpha_p$. This
expression for the longitudinal expansion coefficient is given (but
not derived) in another equivalent form in terms of the Possoin's
Ratio in Refs. \onlinecite{fukao99,bauer00,wallace95} and can be
derived from row 3 of Eq. (53) in Ref.  \onlinecite{christensen07}.
Also note that there is a total lack of standard notation. Bauer
\emph{et al.} use $\alpha_p$ to note the linear expansion coefficient,
which is the quantity often used to express volume expansion of solids. That
is their $\alpha_p$ is 1/3 of our $\alpha_p$. The linear expansion
coefficient is called $\alpha_L$ by Wallace \emph{et
  al.} \cite{wallace95}, while Fukao \emph{et al.} \cite{fukao99} call it
$\alpha_\infty$. The quantity we call the longitudinal expansion
coefficient $\alpha_l$ is denoted $\alpha_{CA}$ (CA for clamped area)
by Bauer, $\alpha_N$ by Wallace and $\alpha_n$ by Fukao (n for
normal).

\section{Relating the measured change in capacitance to
  $\alpha_l$}\label{sec:math}
\subsection{Deriving the relation}
In the measurement we perform a small temperature step $\delta T$ and
 subsequently measure the capacitance $C_m$ as a function of time. 
From the measurements we find the time dependent quantity 
$\frac{1}{C_m}\frac{\Delta C_m}{\Delta T}(t)$. 
In the following section we show that this quantity is proportional to
the expansion coefficient, $\alpha_l(t)$, with a proportionality constant that depends on
$\epsilon_\infty$ and the degree of filling of the capacitor, $f$, but
not on the geometrical capacitance or the distance between the plates.

The starting point is that the only contribution to the high-frequency
dielectric constant, $\epsilon_\infty$, is the atomic polarizability
(Sec. \ref{sec:design}). We moreover use the Lorentz
field \cite{bottcher} from which it follows that dielectric constant is
given by the Clausius-Mossotti relation:
\begin{equation}
\nonumber  
  \frac{\epsilon_{\infty}-1}{\epsilon_{\infty}+2}=\frac{n}{3\epsilon_0}x,
\end{equation}
where $x$ is the polarizability of a single molecule, $n$ is the number
density of molecules, and $\epsilon_0$ is the vacuum permeability.  

Moreover, we assume that we have a parallel plate capacitor which is
partially filled with a dielectric liquid. The degree of filling is denoted by
$f$ and the measured capacitance is given by
\begin{equation}
  \label{eq:1}
  C_{m}=f\epsilon_\infty\frac{A \epsilon_0}{l}+(1-f)\frac{A \epsilon_0}{l}=\[f\epsilon_\infty+(1-f)\] C_g
\end{equation}
where $C_g=\frac{A \epsilon_0}{l}$ is the geometrical capacitance of
the empty capacitor at the given temperature.
 
The derivative with respect to temperature is now given by
\begin{equation}
  \label{eq:cm}
  \diff{C_m}{T}  =\[f\epsilon_\infty+(1-f)\]\diff{C_g}{T}+C_gf\diff{\epsilon_\infty}{T}.
\end{equation}
Here it is assumed that the liquid does not contract radially at the
temperatures (and on the time scale) we consider (see
Sec. \ref{flow}), thus $\textrm{d}f/\textrm{d}T=0$. The next step is
to calculate $\diff{C_g}{T}$ and $\diff{\epsilon_\infty}{T}$ under the
assumption that the area is constant. This was done by
Bauer \cite{bauer00,bauer01}. For completeness we include a detailed
derivation as an Appendix. The result is
\begin{equation}
  \label{eq:3}
  \diff{\epsilon_\infty}{T}= - K(\epsilon_\infty)\alpha_l
\end{equation}
where $K(\epsilon_\infty)$ is given by
$K(\epsilon_\infty)=(\epsilon_\infty-1)(\epsilon_\infty+2)/3$ and
\begin{equation}
  \label{eq:4}
  \diff{C_g}{T}= - C_g\alpha_l\,.
\end{equation}
Inserting Eq. (\ref{eq:3}) and (\ref{eq:4}) in Eq. (\ref{eq:cm}) yields 
\begin{eqnarray}
  \nonumber
  \diff{C_m}{T}=\[f\epsilon_\infty+(1-f)\](- C_g\alpha_l)-C_gf
  K(\epsilon_\infty)\alpha_l\nonumber \\
=-C_g\[f\epsilon_\infty+(1-f)+fK(\epsilon_\infty)\]\alpha_l\,. \nonumber
\end{eqnarray}
Inserting $C_g=C_m/\[f\epsilon_\infty+(1-f)\]$ and dividing by $C_m$  leads to 
\begin{eqnarray}
  \label{eq:7}
\frac{1}{C_m}
\diff{C_m}{T}=-\frac{f\epsilon_\infty+(1-f)+fK(\epsilon_\infty)}{f\epsilon_\infty+(1-f)}\alpha_l\,.
\end{eqnarray}
Isolating finally $\alpha_l$ gives 

\begin{eqnarray}
    \alpha_l=- \frac{f\epsilon_\infty+(1-f)}{f\epsilon_\infty+(1-f)+fK(\epsilon_\infty)}\frac{1}{C_m}\diff{C_m}{T}\nonumber\\
\alpha_l=P(f,\epsilon_\infty))\frac{1}{C_m}\diff{C_m}{T},
\end{eqnarray}
where 
\begin{equation}
  \nonumber
P(f,\epsilon_\infty)=- \frac{f\epsilon_\infty+(1-f)}{f\epsilon_\infty+(1-f)+fK(\epsilon_\infty)}  
\end{equation}

\subsection{The absolute value of $\alpha_l$}
The determination of $\alpha_l$ and also the uncertainties of the
measured value depend on determining correctly the proportionality
constant $P(f,\epsilon_\infty)$. In order to do so we need to
determine the relevant values of $f$ and $\epsilon_\infty$. To find $f$ we use
the expansion coefficient and to find the dielectric constant $\epsilon_\infty$ we use the measured
empty capacitance along with the measured full capacitance.

The high-temperature expansion coefficient is found \cite{ditteExp} to
be 0.7*$10^{-3}$~$K^{-1}$; at low temperatures we find \cite{note} that
it is around 0.5*$10^{-3}$~$K^{-1}$ in the long time limit. We use
0.6*$10^{-3}$~$K^{-1}$ as an average value, and find from this that
the degree of filling is $f=0.95$ if the liquid is assumed to contract
radially down to 213~K where the relaxation time is 100~s. The choice
of expansion coefficient in the range 0.5-0.7*$10^{-3}$~$K^{-1}$ and
final temperatures in the range 210-215~K makes $f$ change with $\pm
1$\%. The effect of changing $f$ within this range leads only to $\pm
0.5$\% changes in $P(f,\epsilon_\infty)$.

Isolating the dielectric constant from Eq. (\ref{eq:1}) gives:
\begin{eqnarray}
  \label{eq:9}
  \epsilon_\infty=\frac{C_m-C_g(1-f)}{fC_g}
\end{eqnarray}
From this it is seen that the uncertainty in $f$ also gives an
uncertainty in $\epsilon_\infty$, and this actually has a greater
impact on the uncertainty of $P$ than the direct effect of the
uncertainty on $f$. Including this effect, the uncertainty in $P$ due
to uncertain degree on filling is still only $\pm 1$\%. 

In order to determine $\epsilon_\infty$ from Eq. (\ref{eq:9}) we need to
know the geometric capacitance, $C_g$. This is found from
measurements on the empty capacitor at the measuring temperature. We
estimate that the uncertainty is $\pm 2$\% on $C_g$. This estimate
is made by comparing measurements made on the capacitor after
assembling it anew. The total uncertainty on $\epsilon_\infty$ is roughly
$\pm 3$\%, which leads to an uncertainty on $P$ of $\pm 2$\%.

Altogether the uncertainty on $P(f,\epsilon_\infty)$ and therefore on
the absolute value of $\alpha_l$ is about $\pm 3$\%. It should be
emphasized that this uncertainty has no effect on the shape or the time
scale of the measured relaxation. This is so as long as we stick to
linear experiments. For larger temperature steps there will be (at least
in principle) some second-order effects making
$P(f,\epsilon_\infty)$ change during the relaxation because of the
change in $\epsilon_\infty$.

In the modelling of the connection between measured change in
capacitance to $\alpha_l$ we have not considered the radial expansion
of the electrode plates. Including this (in the simples possible way)
gives rise to an extra additive term
$\frac{1}{C_m}\frac{\epsilon_0}{l}\diff{A}{T}$ in Eq. (\ref{eq:7}). The
size of this term will be given by the linear expansion coefficient of
the electrodes. They are in this case made of copper, which at the
relevant temperature has a linear expansion of approximately
$15\times10^{-6}$~K$^{-1}$. The total measured change in the capacitance
is about 50-100 times bigger, thus the effect is small. However, the
time dependence is different therefore it could be relevant to include
this effect in the future. Alternatively we also consider shifting to
an electrode material with an even smaller expansion coefficient in
order to avoid the effect all together.

It should be kept in mind that we have used the Lorentz field. This is
an important assumption and the use of an other local field, when
connecting density with the dielectric constant will change the
result. Using the macroscopic Maxwell field, will yield the same
everywhere, except for $K(\epsilon_\infty)$ in Eq. (\ref{eq:3})
which will be given by $K_{Max}(\epsilon_\infty)=(\epsilon_\infty-1)$
instead of the
$K_{Lor}(\epsilon_\infty)=(\epsilon_\infty-1)(\epsilon_\infty+2)/3$. This
leads to a 20~\% increase in $P$ and the calculated numerical value of
$\alpha_l$. Again we stress that using another local field will change
the absolute values, but will not change the time scale or shape of
the measured relaxation.

While none of the above-mentioned things affect the time scale or
the spectral shape of the measured relaxation, the temperature
dependence of $\epsilon_\infty$ could in principle affect the
temperature dependence of the calculated $\alpha_l$. However, this
effect is negligible over the 6 degree range studied in the work and
$P$ will be considered constant.

To summarize, the problems discussed in this section can lead to an
unknown temperature- and frequency-independent scaling of all the
measured $\alpha_l$-values.

\subsection{The shape of the relaxation curve}

In the following we describe the measuring protocol in detail and
describe a correction made on the data. We moreover use this to give
an estimate of the uncertainty on the shape of the relaxation curves
reported. 

\begin{figure}
  \centering
  \includegraphics[width=0.85\linewidth]{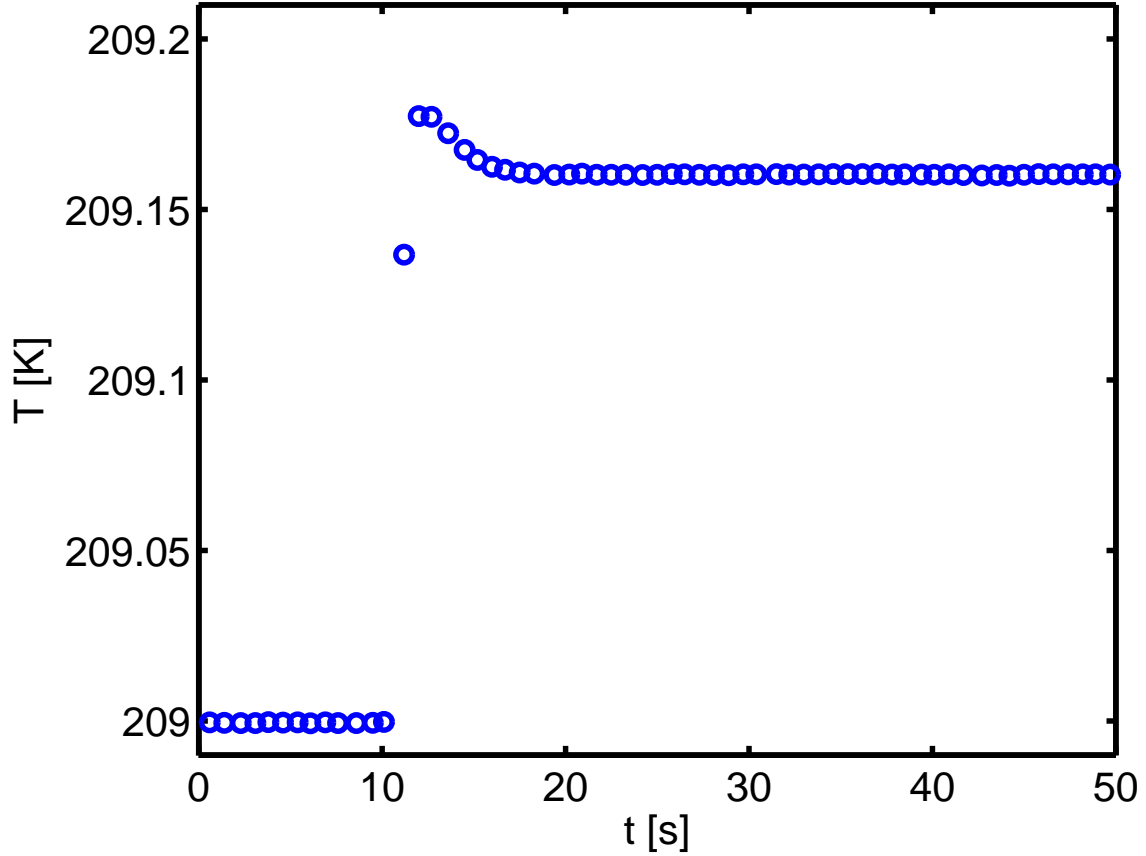}
  \caption{Zoom on the temperature monitored in the NTC-bead in the
    lower capacitor plate during the first 40 seconds of at temperature step.}\label{fig:Tstep}
\end{figure}

A main issue is, of course, the first part of the measuring curve where
the temperature gets in equilibrium. Fig. (\ref{fig:Tstep})
shows details of a single temperature step. It is clearly seen how the target
temperature is achieved within less than 10~s, corresponding to a
characteristic time of 2~s. 

\begin{figure}
  \centering
  \includegraphics[width=0.3\linewidth]{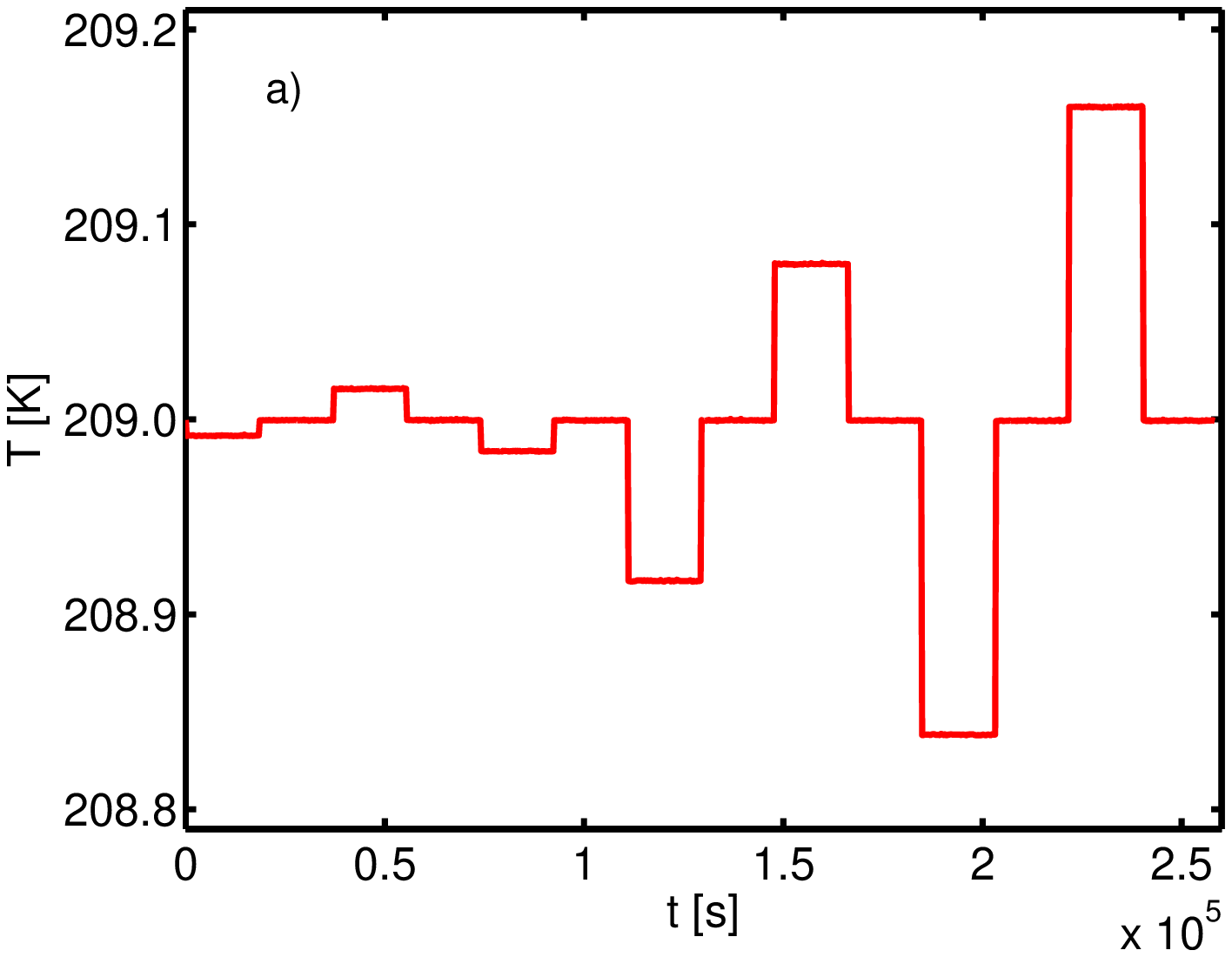}
\includegraphics[width=0.3\linewidth]{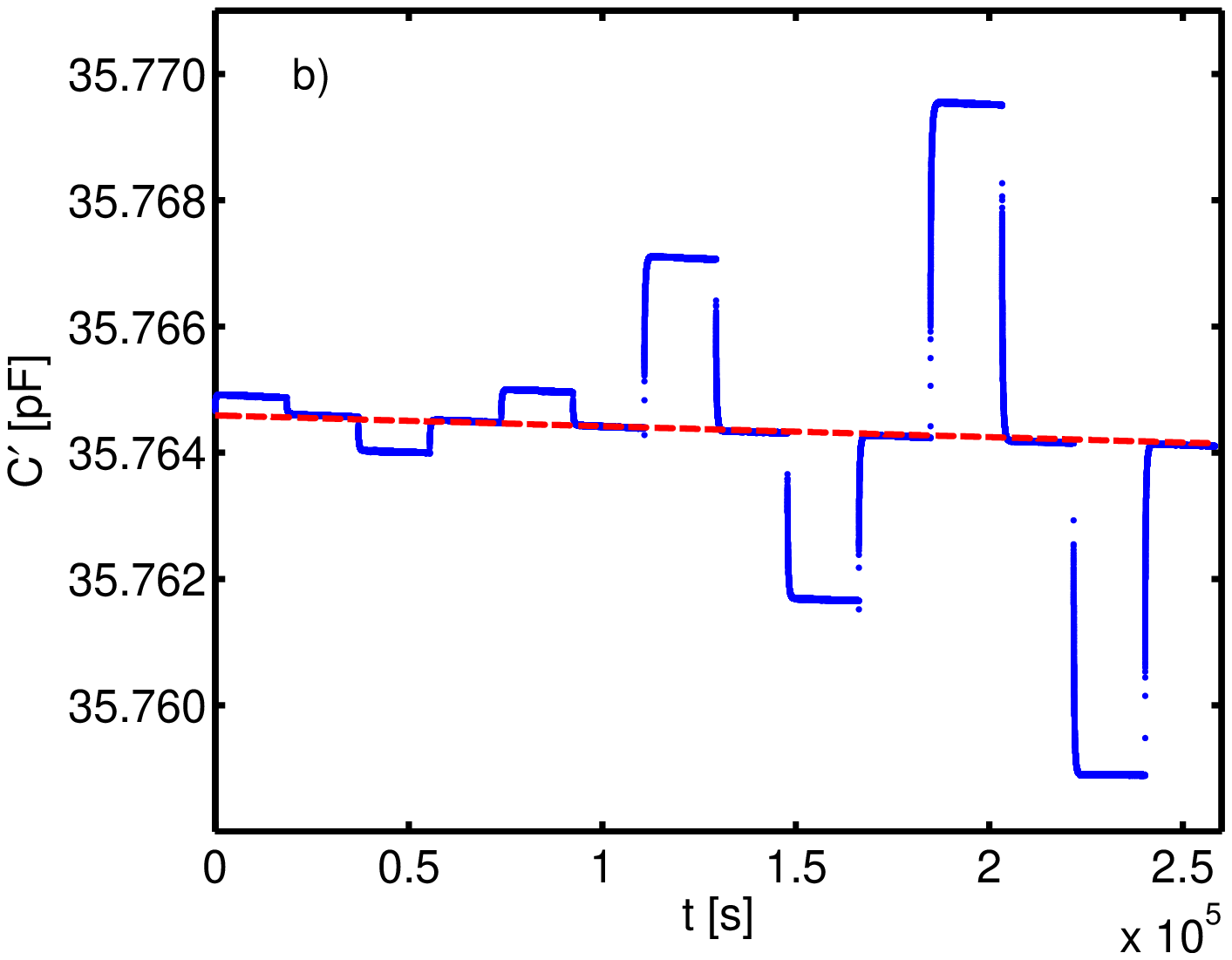}
\includegraphics[width=0.3\linewidth]{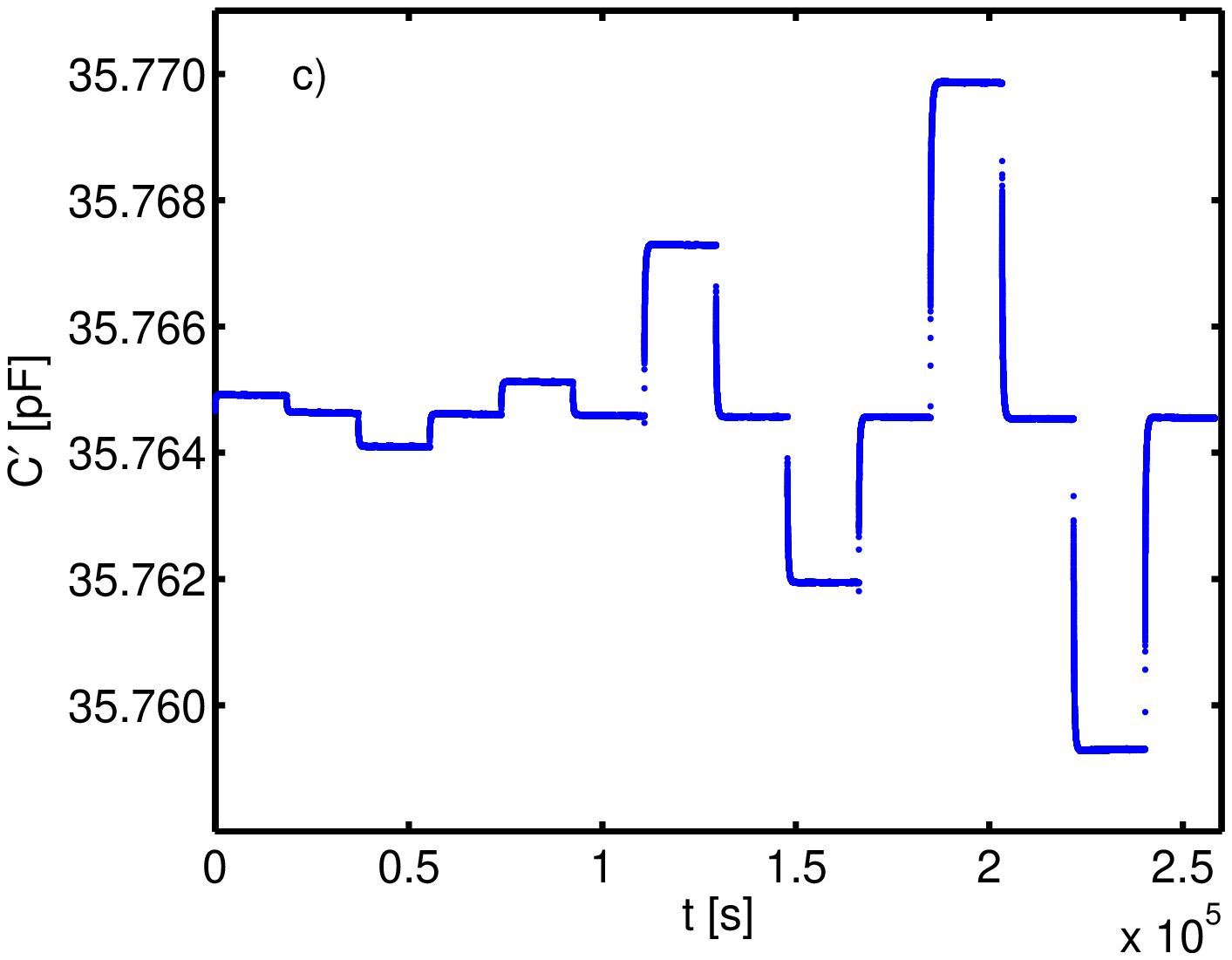}  
\caption{a) Example of temperature protocol at one temperature. A
    series of up and down jumps are made at the same reference
    temperature with different amplitude. The temperatures shown are
    those measured with the NTC-bead in the lower capacitor plate. Notice that the smallest
    jumps are 0.01~K.  b) The measured capacitance (blue points). Notice that the
    relative changes in capacitance ($\textrm{d}C/C$) for the small
    jumps is less than 10$^{-4}$ and can still be measured
    precisely. There is a long time drift in the measured capacitance,
    the dashed line illustrates this background drift and this slope
    is subtracted of the data before further
    treatment.\\
c) The measured capacitance after subtraction of the
    drift.
}\label{fig:rawCsteps}
\end{figure}

Fig. (\ref{fig:rawCsteps}) a) shows a typical set of temperature steps: a
series of up and down jumps are made at the same temperature, with
variable amplitude. 

Fig. (\ref{fig:rawCsteps}) b) shows the raw measured capacitance
corresponding to the temperature steps in Fig. (\ref{fig:rawCsteps}) a). Two
things are worth noticing. First we see the expected rise in
capacitance when temperature is decreased. Secondly, we see a long time
drift of the equilibrium level.
At low temperatures where the liquid cannot contract radially it
contracts vertically. Comparing measurements on the empty capacitor
with liquid filled measurements we estimate that the expansion
coeficient of the liquid is roughly 10 times larger than that of the
Kapton spacers. This means that the liquid compresses the
Kapton. However, on very long times it will be the Kapton which
dominates (becuase the liquid flows) and the Kapton will therefore
slowly relax and press the electrodes apart. We believe that this
effect is what leads to the long time drift seen in
Fig. (\ref{fig:rawCsteps}) b).
\begin{figure}
  \centering
  \includegraphics[width=0.85\linewidth]{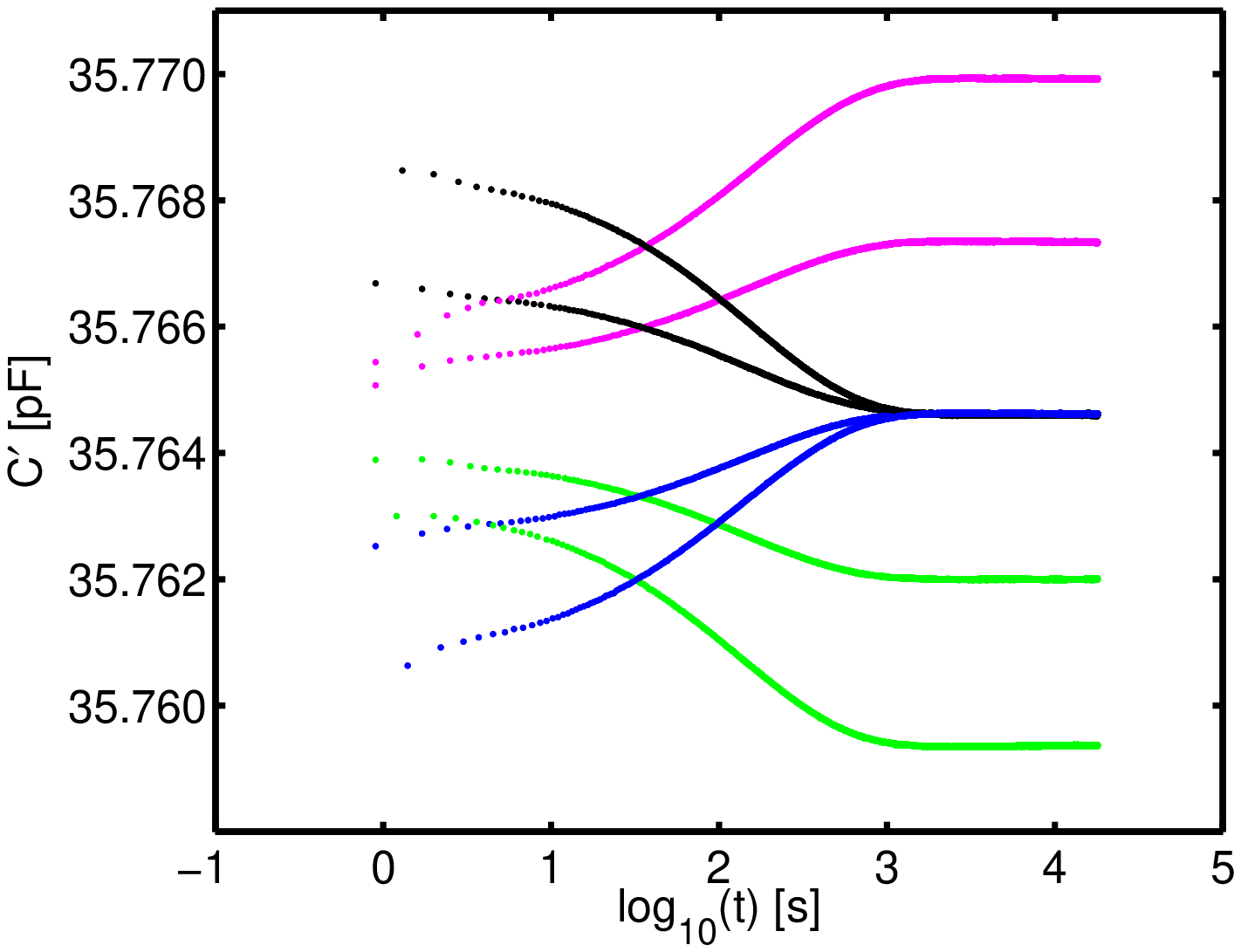}
  \caption{The corrected measured capacitance shown on a logarithmic
    time scale with the temperature change as starting time. The
    temperature steps with the same final temperature (shown in black
    and blue) all have the same final value of the 
    capacitance. The data shown here correspond to the last 8-steps in
  Fig. (\ref{fig:rawCsteps})}\label{fig:logcorrCsteps}
\end{figure}
The drift is subtracted before treating the data, as illustrated in
Fig. (\ref{fig:rawCsteps}) c) and Fig. (\ref{fig:logcorrCsteps}).

\begin{figure}
  \centering
  \includegraphics[width=0.85\linewidth]{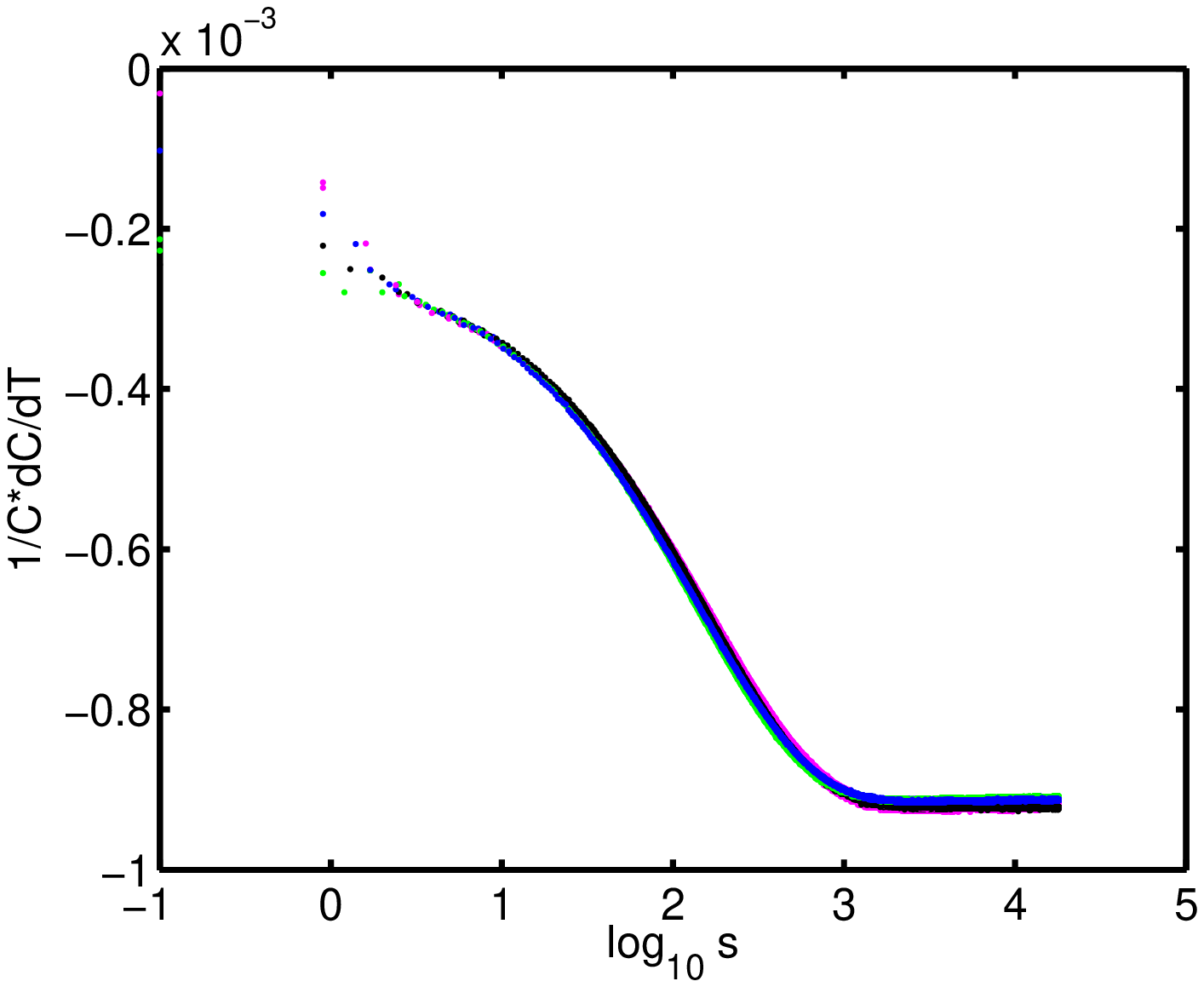}
  \caption{The relative change in C divided
    by the size of the temperature step. The data shown are the same
    data as in Fig. (\ref{fig:logcorrCsteps}) and the colors used for each
    curve is the same. All the curves superpose, which demonstrates
    that the experiment is linear. Moreover it demonstrates that the
    subtraction of the background drift is
    successful.}\label{fig:deltaC}
\end{figure}

We make both up jumps and down jumps in temperature and the subtraction
of the drift has an opposite effect on the two. We can therefore check
that the subtraction is made correctly by comparing up jumps and down
jumps. This is illustrated in Fig. (\ref{fig:deltaC}). The
superposition  of data obtained in up and down jumps also demonstrates
that the experiment is linear and gives a
general estimate of how precise the determination of the curve shape
is.

The comparison of up and down jumps moreover serves to guarantee that
the steps are linear. The relaxation time is strongly temperature
dependent when the liquid is close to the glass transition, and the
steps therefore have to be very small in order to maintain linear
behavior. Smaller steps can be made as well, and the shape of the
relaxation is maintained, but the curve starts to get noisy because the
signal is very small. When we make larger temperature steps, we begin 
to get typical non-linear aging behavior. That is, the relaxation is
slower for down jumps than for up jumps when the final temperature is
the same.
The setup is actually well suited for
nonlinear experiments also; because of the extremely high resolution
of the measured quantity we get very well-defined curves and can
clearly see the nonlinear behavior already at steps of 1 degree. We
plan to use the setup for these types of studies, as well, but focus in this
paper on the linear results. 

\section{Data}

\begin{figure}
  \centering
  \includegraphics[width=\linewidth]{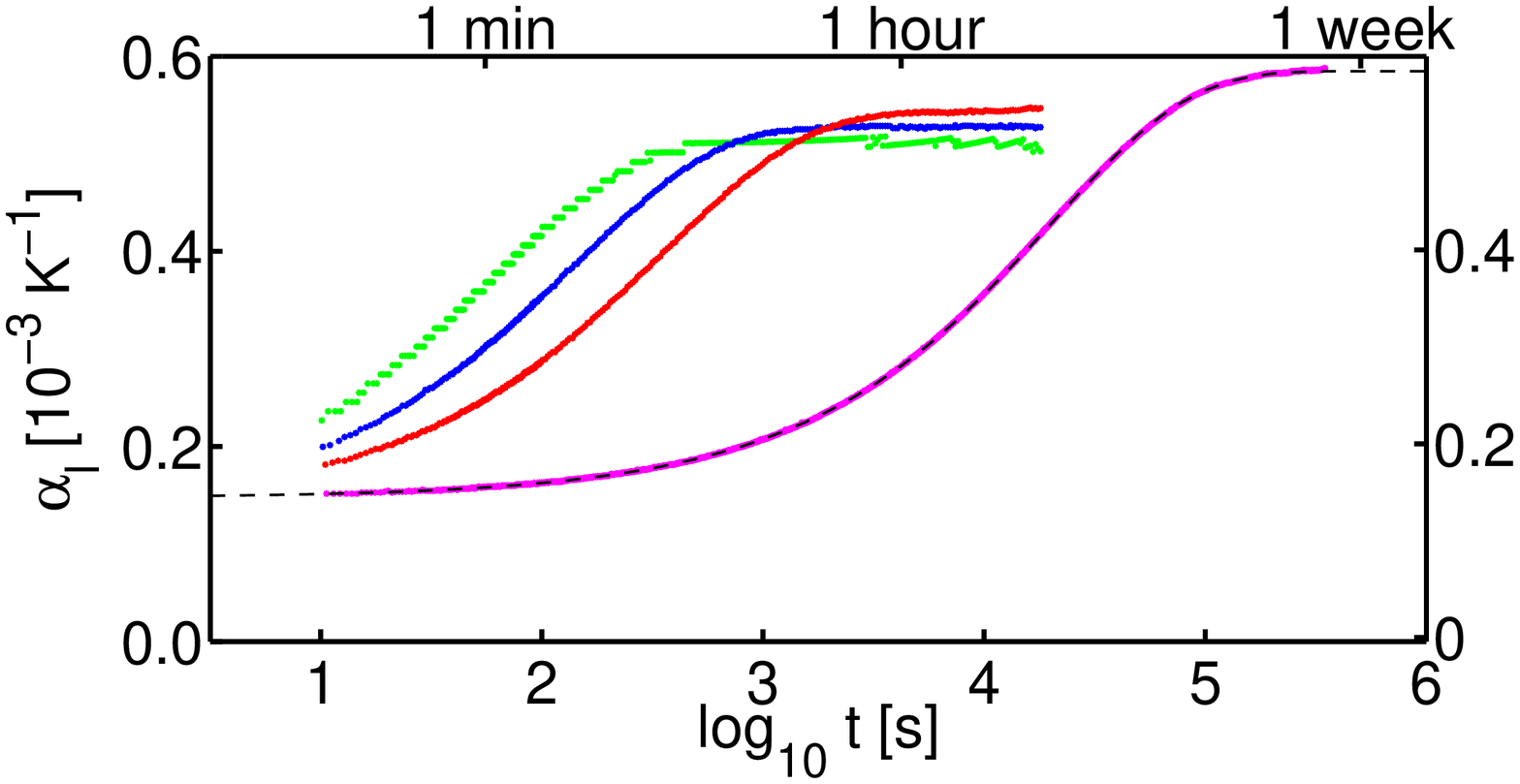}
  \caption{The measured time dependent expansion coefficient of
    Tetramethyltetraphenyltrisiloxane at T= 205~K, 209~K, 210~K and
    211~K. At the lowest temperature we also show a fit to the
    modified stretched exponential \cite{saglanmak}.}\label{fig:rawdat}
\end{figure}

Fig. (\ref{fig:rawdat}) shows the expansion coefficient as a function
of time at four different temperatures. The data are shown for steps made
with $\approx$0.1~K, except the data at 211~K which are taken with a
temperature step of $\approx$0.01~K. This is why there is more noise
on this dataset.    

\begin{figure}
  \centering
  \includegraphics[width=\linewidth]{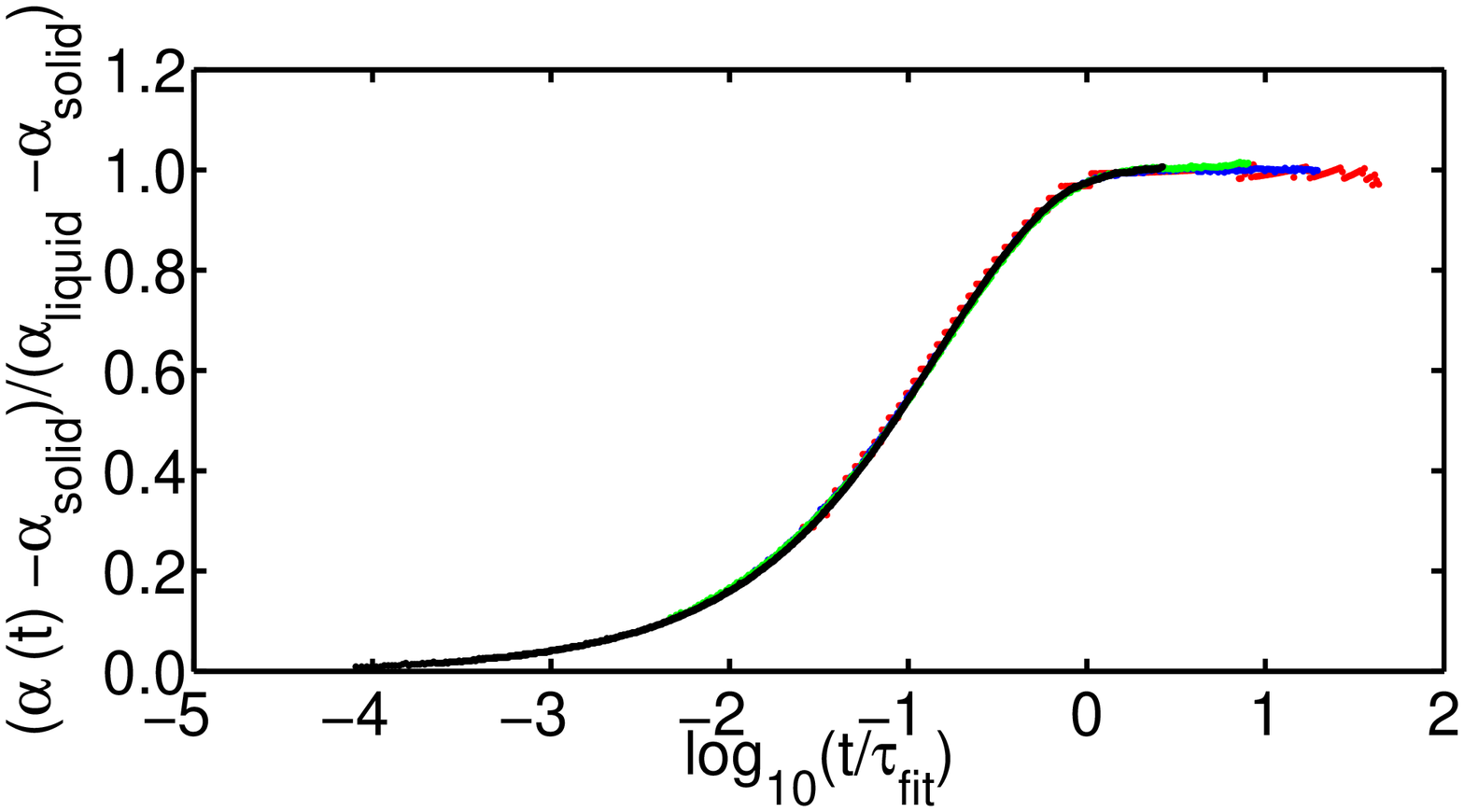}
  \caption{The data from Fig. (\ref{fig:rawdat}) normalized and
    plotted versus time scaled with the relaxation time (as defined
    from the fit to the modified stretched exponential). The figure
    demonstrates that the data obey time temperature superposition
    (TTS). }\label{fig:tts}
\end{figure}

Fig. (\ref{fig:tts}) shows all the data from Fig. (\ref{fig:rawdat})
normalized and superimposed. This illustrates that the measured
relaxation obeys time-temperature-superposition (TTS) within the
studied (relatively narrow) temperature range. The fit shown in figure
\ref{fig:rawdat} is a fit to the superimposed curve obtain from the
datasets at T=205~K and T=211~K.

\begin{figure}
  \centering
  \includegraphics[width=\linewidth]{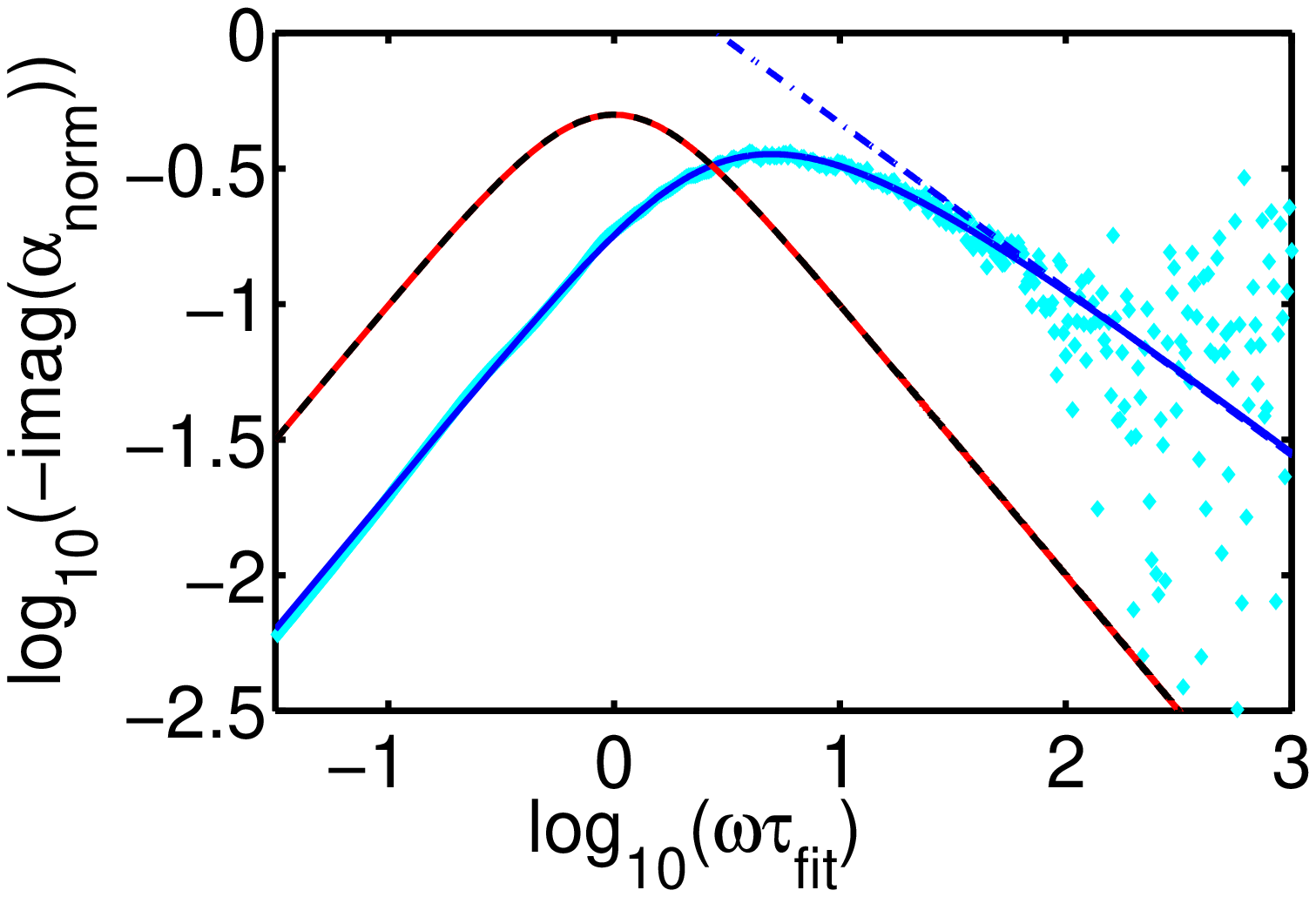}
  \caption{Illustration of the Laplacetransform. Cyan Diamonds:
    Laplace Transform of the normalized data (found by brute force
    nummerical integration of the measured points). Blue:
    Laplacetransform of the normalized fit (see text for
    details). Red: Exponential relaxation which has been transformed
    using the same algorithm as the used for the fit. Black dashed :
    analytical Laplace transform of Exponential relaxation. Blue
    dashed-dotted
    line: power law, which corresponds to the exponent of the
    fit.}\label{fig:LapIllu}
\end{figure}

The function used to fit the date is a modified stretched
exponential \cite{saglanmak} given by:
\begin{equation}
  \label{eq:25}
\alpha(t) =\alpha_{\infty}+\Delta \alpha\left(1-\exp\left[-k\left(\frac{t}{\tau}\right)^{\bkww}-\frac{t}{\tau} \right]\right).
\end{equation}
In the fit to data we get $\bkww=0.6$ and $k=2.6$. The quality of the fit is so good
that we have used it as an interpolation of the data and used it to
calculate the frequency-domain response, which is given by the
transformation in Eq. (\ref{Rw}). The transformation is made by making
a discrete Fourier transform  (using matlabs FFT-procedure) on the fit of the normalized curve
evaluated in a number of points.  The transformed normalized curve is
shown in Fig. (\ref{fig:LapIllu}). Here we also show an Exponential
relaxation which has been transformed using the same algorithm along
with the analytical Laplace transform. Moreover, the high frequency
power law, which corresponds to the exponent of the fit, is also
shown. 

\begin{figure}
  \centering
  \includegraphics[width=\linewidth]{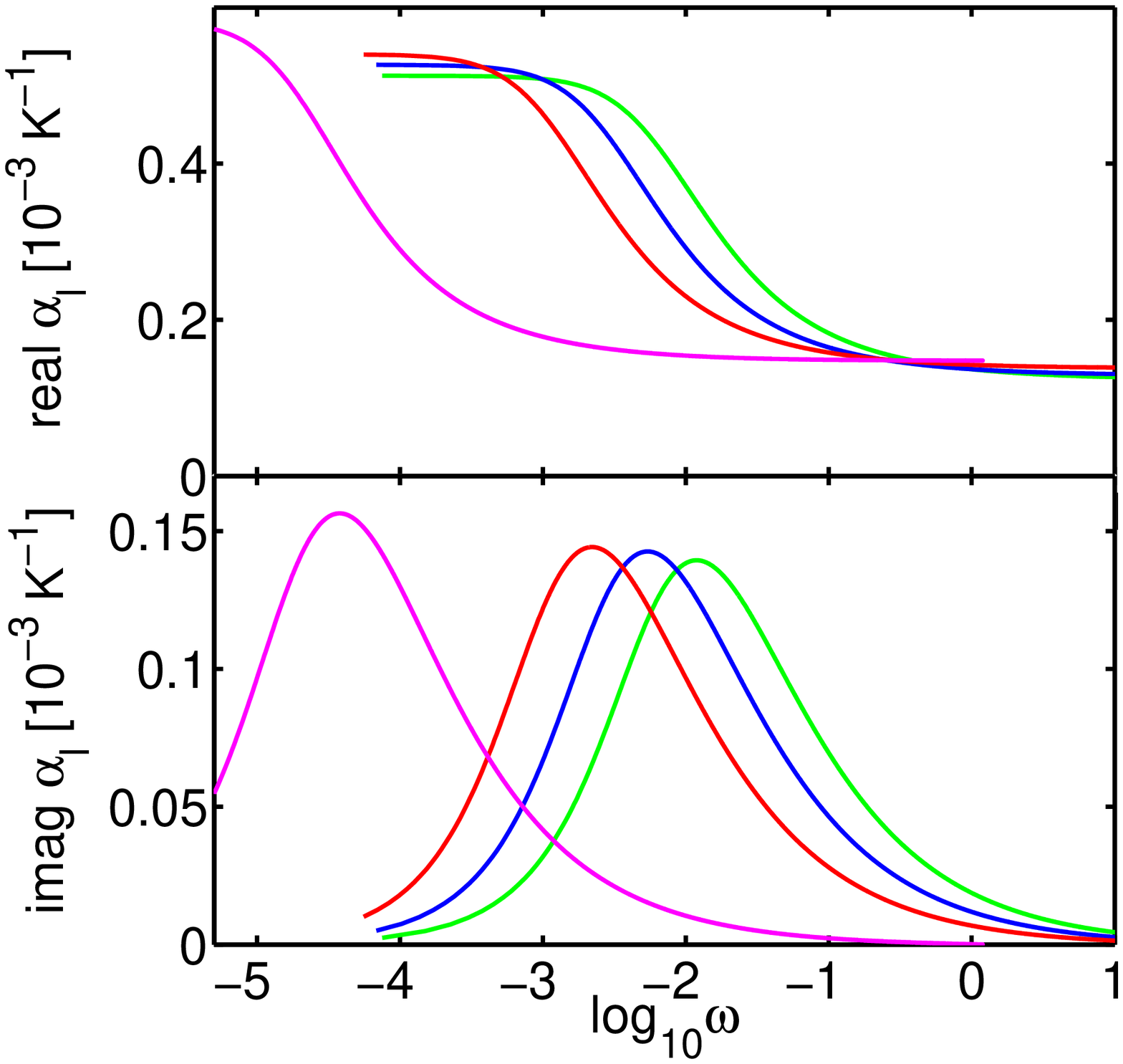}
  \caption{Laplace transformed fits (see the text for details). The
    curves are shown a in the dynamical range
    that roughly corresponds to the measurement.}\label{fig:FreqDom}
\end{figure}

In Fig. (\ref{fig:FreqDom}) we show the Laplace transformed fit
rescaled with amplitudes and time scales in order to show the
temperature dependence of the frequency dependent thermal expansion
coefficient. 

\section{Summary and outlook}
We have presented a technique for measuring the dynamical
expansion coefficient $\alpha(t)$ for a glass-forming liquid in the
ultraviscous range. The experiment is performed on a setup which
follows the capacitative principle suggested by Bauer \emph{et al.} \cite{bauer00}. The
dynamical range has been extended from 1.5 decade to more than four
decades by making time-domain experiments,
and by making very small and fast temperature steps. The modelling of
the experiment has moreover been developed.  Data is presented on the
molecular glass-former tetramethyl tetraphenyl trisiloxane (DC704). This
data set is to the best of our knowledge the first data on the dynamical expansion
coefficient of a molecular liquid. 

The technique presented in this paper is based on a principle where
the sample is placed in a parallel plate capacitor such that the
sample maintains the spacing between the plates. Thus a change in
sample volume in response to a temperature change leads to a change of
the capacitance. The advantage of this technique is that capacitances
can be measured with very high precision, and the small density
changes associated with linear experiments can therefore be determined
reliably. One limitation of the technique is that it only works on
timescales larger than 10~s, this could possibly be overcome by
smaller samples and thereby faster temperature control. A more
intrinsic limitation is that the technique only works for samples with
very small dipole moment. For samples with large dipole moment we
therefore need a complementary technique.

The measurements of the thermal expansivity is part of a general
ambition in the ``Glass and Time'' group to measure different response
function of viscous liquids. A unique feature of our techniques is
that the all fit into the same type of cryostat \cite{igarashi08},
ensuring that the absolute temperature of the liquid is the same for
all measurements. The thermal expansion measurements described in this
paper are thus performed in the same cryostat as our shear mechanical
spectroscopy \cite{christensen95}, bulk mechanical spectroscopy
\cite{christensen94}, specific heat spectroscopy \cite{jakobsen10} and
dielectric spectroscopy \cite{igarashi08b}. The properties of liquids
close to the glass transition are extremely temperature dependent,
and small differences in the temperature calibration can lead to
rather large differences in the results. Measuring different
response functions at the exact same conditions therefore makes it possible to
analyze new aspects of the viscous slowing down and the glass
transition. In recent papers we used this to compare time
scales of all the different response functions \cite{jakobsen11}, to
relate linear response to density scaling and to determine the linear
Prigogine Defay ratio \cite{gundermann11}.

\section{Acknowledgments}
The center for viscous liquid dynamics ``Glass and Time'' is sponsored
by the Danish National Research Foundation (DNRF).  Ib Høst
Pedersen, Torben Rasmussen, Ebbe Larsen and Preben Larsen are thanked
for their contribution to development of the temperature control and
the measuring cell. Niels Boye Olsen is thanked for sharing his
experience and ideas. Tina Hecksher and Bo Jakobsen are thanked for
fruitful discussions.

\section*{Appendix I }
The relation  between the temperature derivative of the dielectric
constant and of the geometrical capacitance with the longitudinal
expansion coefficient was derived by Bauer \cite{bauer00,bauer01}. For completeness we
include a detailed derivation in this Appendix. 

The longitudinal expansion coefficient is defined by 
\begin{equation}
  \label{eq:12}
  \alpha_l=\frac{1}{l}\(\parti{l}{T}\)_A
\end{equation}
We start with the temperature derivative of $\epsilon_\infty$, which
in this situation is given by:
\begin{equation}
  \label{eq:11}
  \(\parti{\epsilon_\infty}{T}\)=\(\parti{\epsilon_\infty}{l}\)_A\(\parti{l}{T}\)_A
\end{equation}
so we need and expression for the first term,
$\(\parti{\epsilon_\infty}{l}\)_A$. The Clausius-Mossotti relation gives
\begin{equation}
  \label{eq:13}
  \frac{xN}{3\epsilon_0 A l}=\frac{(\epsilon_{\infty}-1)}{(\epsilon_{\infty}+2)}
\end{equation}
where $N$ is the total number of molecules, $A$ is the area and $l$ is
the thickness such that $N/(Al)$ is the number density of molecules
and $x$ the microscopic polarizability of the molecule.

We rewrite this to get
\begin{equation}
  \nonumber
  \epsilon_{\infty}=\frac{xN}{3\epsilon_0 A l}\left(\epsilon_{\infty}+2\right)+1
\end{equation}
and take the derivative with respect to $l$ at constant $A$
\begin{equation}
  \nonumber
  \(\parti{\epsilon_\infty}{l}\)_A=\frac{xN}{3\epsilon_0 A l}\(\parti{\epsilon_\infty}{l}\)_A-(\epsilon_\infty+2)\frac{xN}{3\epsilon_0 A l^2}
\end{equation}
which by reinserting Eq. (\ref{eq:13}) gives
\begin{equation}
  \nonumber
  \(\parti{\epsilon_\infty}{l}\)_A= \frac{(\epsilon_{\infty}-1)}{(\epsilon_{\infty}+2)}
  \(\parti{\epsilon_\infty}{l}\)_A-(\epsilon_\infty+2)\frac{1}{l} \frac{(\epsilon_{\infty}-1)}{(\epsilon_{\infty}+2)}.
\end{equation}
We now isolate $\(\parti{\epsilon_\infty}{l}\)_A$ in this expression and get 
\begin{equation}
  \nonumber
  \(\parti{\epsilon_\infty}{l}\)_A=-\frac{1}{l}\frac{(\epsilon_\infty-1
)(\epsilon_\infty+2)}{3}
\end{equation}
inserting this in Eq. (\ref{eq:11})
\begin{equation}
  \nonumber
  \(\parti{\epsilon_\infty}{T}\)_T=-\frac{(\epsilon_\infty-1
)(\epsilon_\infty+2)}{3} \frac{1}{l}\(\parti{l}{T}\)_A
\end{equation}
which when comparing to the definition of the longitudinal expansion
coefficient in Eq. (\ref{eq:12}) can be rewritten as 
\begin{equation}
  \nonumber
  \(\parti{\epsilon_\infty}{T}\)_T=-\frac{(\epsilon_\infty-1
)(\epsilon_\infty+2)}{3} \alpha_l=-K(\epsilon_\infty)\alpha_l
\end{equation}
where the last equality comes from defining  $K(\epsilon_\infty)=\frac{(\epsilon_\infty-1
)(\epsilon_\infty+2)}{3}$

Now we move on to the temperature derivative of the geometrical
capacitance, $C_g$, which
in this situation is given by:
\begin{equation}
  \label{eq:23}
  \(\parti{C_g}{T}\)=\(\parti{C_g}{l}\)_A\(\parti{l}{T}\)_A
\end{equation}
The geometrical capacitance itself is given by 
\begin{equation}
  \nonumber
  C_g=\frac{A\epsilon_0}{l}
\end{equation}
giving
\begin{equation}
  \nonumber
  \(\parti{C_g}{l}\)_A=-\frac{A\epsilon_0}{l^2}=-\frac{1}{l}C_g
\end{equation}
which when inserted in Eq. (\ref{eq:23}) and combined with the definition of
the longitudinal expansion coefficient gives 
\begin{equation}
  \label{eq:22}
  \(\parti{C_g}{T}\)_A=-C_g\alpha_l.
\end{equation}

\section{Appendix II :  FD-theorem and the expansion coefficient}
This appendix gives the formal definition of the dynamic expansion
coefficient, including how it relates to fluctuations and how the
frequency-domain response is related to the measured time-domain
response. This is and extension of the presentation in
Ref. \onlinecite{bauer00}. However, Ref. \onlinecite{bauer00} contains a typo as
well some definitions which are not precise regarding the absolute
levels of the response functions. The precise definitions are
important for our use of the data in Ref. \onlinecite{gundermann11}.

  The measured response to an external field, whether in the
time domain or in the frequency domain, is directly related to the
equilibrium thermal fluctuations of the system. This is expressed
formally through the fluctuation dissipation theorem (FDT), which
expressed in the time domain is\cite{doi,nielsen96}:
\begin{equation}\label{timeFDT}
 \dfrac{dR(t)}{dt}=-\dfrac{1}{k_BT}\dfrac{d}{dt}\langle \Delta A(t)\Delta B(0)\rangle.
\end{equation}
Here $R(t)$ is the response function (see Sec. (\ref{sec:response})
for a definition) and sharp brackets refers to ensemble averages. $A$
is the measured physical quantity (that is the output $O(t)$ in
Sec. (\ref{sec:response})) and $B$ is conjugated to the applied
input/field, which is called $I(t)$ in Sec.
(\ref{sec:response}). The function $\langle \Delta A(t)\Delta
B(0)\rangle$ is the correlation function, which in the simple case
where $A=B$ reduces to the auto correlation function.

Integrating on both sides of Eq. (\ref{timeFDT}) and inserting
$R(t=0)=0$ gives the time-domain response function :
\begin{eqnarray}\label{timeFDT2}
  \int_0^t \dfrac{dR(t')}{dt}dt'&=&-\int^t_0 \dfrac{1}{k_BT}\dfrac{d}{dt'}\langle \Delta A(t') \Delta B(0)\rangle dt'\nonumber \\
  R(t)&=&\dfrac{1}{k_BT}(\langle \Delta A(0)\Delta B(0)\rangle - \langle \Delta A(t)\Delta B(0)\rangle).
\end{eqnarray}
from which it is seen that $R(t=0)=0$ as it should be. 
The frequency-domain response function, is given by the Laplace transform of $R(t)$ times $i\omega$:
\begin{equation}\label{Rw}
 R(\omega)=i\omega \int^\infty_0R(t')e^{-i\omega t'}dt'.
\end{equation}
Combining this with Eq. (\ref{timeFDT2}) gives the FDT in the frequency
domain:
\begin{eqnarray}\label{freqFDT}
R(\omega)&=&-\dfrac{i\omega}{k_BT}\int^\infty_0 \langle \Delta A(t)\Delta
B(0)\rangle -\langle \Delta A(0)\Delta B(0)\rangle e^{-i\omega t}dt\nonumber \\
&=&\dfrac{1}{k_BT}\langle \Delta A(0) \Delta
B(0)\rangle-\dfrac{i\omega}{k_BT}\int^\infty_0\langle \Delta
A(t)\Delta B(0)\rangle e^{-i\omega t}dt.
\end{eqnarray}

Considering now a linear experiment where a small temperature step
$\delta T$ is applied to a system at constant pressure at $t=0$. Its volume
response is subsequently measured as a function of time:
\begin{equation}
 \delta V(t)=R(t-t') \delta T(t'),
\end{equation}
then the response function $R(t)$ is given by
$R(t)=\frac{\delta V(t)}{\delta T}$ (see Sec. \ref{sec:response}
for more details on the linear response formalism). The time-dependent
isobaric expansion coefficient is defined by
\begin{eqnarray}
 \alpha_p(t)&=&\dfrac{1}{V}\frac{\delta V(t)}{\delta T}\nonumber \\
&=&\dfrac{R(t)}{V}
\end{eqnarray}
In terms of the FDT (Eq. (\ref{timeFDT2})), the relevant
fluctuations for $\alpha_p(t)$ are volume and entropy, and the
expansion coefficient can therefore be expressed in the following way:
\begin{eqnarray}
   \alpha_p(t)&=&\dfrac{1}{Vk_bT}\left(\langle \Delta V(0)\Delta S(0)\rangle-\langle \Delta V(t)\Delta S(0)\rangle\right).
\end{eqnarray}
The frequency-domain response function $\alpha_p(\omega)$ is then (from Eq. (\ref{freqFDT}))
\begin{equation}
\alpha_P(\omega)=\dfrac{1}{Vk_BT}\langle \Delta V(0)\Delta
S(0)\rangle-\dfrac{i\omega}{Vk_BT}\int^\infty_0 \langle \Delta
V(t)\Delta S(0)\rangle e^{-i\omega t}dt.
\end{equation}

\pagebreak


\begin{thebibliography}{38}
\expandafter\ifx\csname natexlab\endcsname\relax\def\natexlab#1{#1}\fi
\expandafter\ifx\csname bibnamefont\endcsname\relax
  \def\bibnamefont#1{#1}\fi
\expandafter\ifx\csname bibfnamefont\endcsname\relax
  \def\bibfnamefont#1{#1}\fi
\expandafter\ifx\csname citenamefont\endcsname\relax
  \def\citenamefont#1{#1}\fi
\expandafter\ifx\csname url\endcsname\relax
  \def\url#1{\texttt{#1}}\fi
\expandafter\ifx\csname urlprefix\endcsname\relax\def\urlprefix{URL }\fi
\providecommand{\bibinfo}[2]{#2}
\providecommand{\eprint}[2][]{\url{#2}}

\bibitem[{\citenamefont{Angell et~al.}(2000)\citenamefont{Angell, Ngai,
  McKenna, McMillan, and Martin}}]{angell00}
\bibinfo{author}{\bibfnamefont{C.~A.} \bibnamefont{Angell}},
  \bibinfo{author}{\bibfnamefont{K.~L.} \bibnamefont{Ngai}},
  \bibinfo{author}{\bibfnamefont{G.~B.} \bibnamefont{McKenna}},
  \bibinfo{author}{\bibfnamefont{P.~F.} \bibnamefont{McMillan}},
  \bibnamefont{and} \bibinfo{author}{\bibfnamefont{S.~W.}
  \bibnamefont{Martin}}, \bibinfo{journal}{J. Appl. Phys.}
  \textbf{\bibinfo{volume}{88}}, \bibinfo{pages}{3113} (\bibinfo{year}{2000}).

\bibitem[{\citenamefont{Dyre}(2006)}]{dyre06}
\bibinfo{author}{\bibfnamefont{J.~C.} \bibnamefont{Dyre}},
  \bibinfo{journal}{Rev. Mod. Phys.} \textbf{\bibinfo{volume}{78}},
  \bibinfo{pages}{953} (\bibinfo{year}{2006}).

\bibitem[{\citenamefont{Davies and Jones}(1953)}]{davies53}
\bibinfo{author}{\bibfnamefont{R.~O.} \bibnamefont{Davies}} \bibnamefont{and}
  \bibinfo{author}{\bibfnamefont{G.~O.} \bibnamefont{Jones}},
  \bibinfo{journal}{Adv. Phys.} \textbf{\bibinfo{volume}{2}},
  \bibinfo{pages}{370} (\bibinfo{year}{1953}).

\bibitem[{\citenamefont{Prigogine and Defay}(1954)}]{prigogine54}
\bibinfo{author}{\bibfnamefont{I.}~\bibnamefont{Prigogine}} \bibnamefont{and}
  \bibinfo{author}{\bibfnamefont{R.}~\bibnamefont{Defay}},
  \emph{\bibinfo{title}{Chemical {T}hermodynamics}}
  (\bibinfo{publisher}{Longmans, London}, \bibinfo{year}{1954}).

\bibitem[{\citenamefont{Goldstein}(1963)}]{goldstein63}
\bibinfo{author}{\bibfnamefont{M.}~\bibnamefont{Goldstein}},
  \bibinfo{journal}{J. Chem. Phys.} \textbf{\bibinfo{volume}{39}},
  \bibinfo{pages}{3369} (\bibinfo{year}{1963}).

\bibitem[{\citenamefont{Moynihan and Gupta}(1978)}]{moynihan78}
\bibinfo{author}{\bibfnamefont{C.~T.} \bibnamefont{Moynihan}} \bibnamefont{and}
  \bibinfo{author}{\bibfnamefont{P.~K.} \bibnamefont{Gupta}},
  \bibinfo{journal}{J. Non-Cryst. Solids} \textbf{\bibinfo{volume}{29}},
  \bibinfo{pages}{143} (\bibinfo{year}{1978}).

\bibitem[{\citenamefont{Ellegaard et~al.}(2007)\citenamefont{Ellegaard,
  Christensen, Christiansen, Olsen, Pedersen, Schr\o{}der, and
  Dyre}}]{ellegaard07}
\bibinfo{author}{\bibfnamefont{N.~L.} \bibnamefont{Ellegaard}},
  \bibinfo{author}{\bibfnamefont{T.}~\bibnamefont{Christensen}},
  \bibinfo{author}{\bibfnamefont{P.~V.} \bibnamefont{Christiansen}},
  \bibinfo{author}{\bibfnamefont{N.~B.} \bibnamefont{Olsen}},
  \bibinfo{author}{\bibfnamefont{U.~R.} \bibnamefont{Pedersen}},
  \bibinfo{author}{\bibfnamefont{T.~B.} \bibnamefont{Schr\o{}der}},
  \bibnamefont{and} \bibinfo{author}{\bibfnamefont{J.~C.} \bibnamefont{Dyre}},
  \bibinfo{journal}{J. Chem. Phys.} \textbf{\bibinfo{volume}{126}},
  \bibinfo{pages}{074502} (\bibinfo{year}{2007}).

\bibitem[{\citenamefont{Angell}(1991)}]{Angell1991}
\bibinfo{author}{\bibfnamefont{C.~A.} \bibnamefont{Angell}},
  \bibinfo{journal}{J. Non-Cryst. Solids} \textbf{\bibinfo{volume}{131-133}},
  \bibinfo{pages}{13 } (\bibinfo{year}{1991}).

\bibitem[{\citenamefont{Kovacs}(1958)}]{kovacs58}
\bibinfo{author}{\bibfnamefont{A.~J.} \bibnamefont{Kovacs}},
  \bibinfo{journal}{J. Pol. Sci.} \textbf{\bibinfo{volume}{30}},
  \bibinfo{pages}{131} (\bibinfo{year}{1958}).

\bibitem[{\citenamefont{Greiner and Schwarzl}(1989)}]{greiner89}
\bibinfo{author}{\bibfnamefont{R.}~\bibnamefont{Greiner}} \bibnamefont{and}
  \bibinfo{author}{\bibfnamefont{F.}~\bibnamefont{Schwarzl}},
  \bibinfo{journal}{Coll. and Pol. Sci.} \textbf{\bibinfo{volume}{267}},
  \bibinfo{pages}{39} (\bibinfo{year}{1989}).

\bibitem[{\citenamefont{Kolla and Simon}(2005)}]{kolla05}
\bibinfo{author}{\bibfnamefont{S.}~\bibnamefont{Kolla}} \bibnamefont{and}
  \bibinfo{author}{\bibfnamefont{S.~L.} \bibnamefont{Simon}},
  \bibinfo{journal}{Polymer} \textbf{\bibinfo{volume}{46}},
  \bibinfo{pages}{733} (\bibinfo{year}{2005}).

\bibitem[{\citenamefont{Svoboda et~al.}(2006)\citenamefont{Svoboda, Pustkova,
  and Malek}}]{svoboda06}
\bibinfo{author}{\bibfnamefont{R.}~\bibnamefont{Svoboda}},
  \bibinfo{author}{\bibfnamefont{P.}~\bibnamefont{Pustkova}}, \bibnamefont{and}
  \bibinfo{author}{\bibfnamefont{J.}~\bibnamefont{Malek}}, \bibinfo{journal}{J.
  Non-Cryst. Solids} \textbf{\bibinfo{volume}{352}}, \bibinfo{pages}{42}
  (\bibinfo{year}{2006}).

\bibitem[{\citenamefont{Bauer et~al.}(2000{\natexlab{a}})\citenamefont{Bauer,
  B\"ohmer, Moreno-Flores, Richert, Sillescu, and Neher}}]{bauer00}
\bibinfo{author}{\bibfnamefont{C.}~\bibnamefont{Bauer}},
  \bibinfo{author}{\bibfnamefont{R.}~\bibnamefont{B\"ohmer}},
  \bibinfo{author}{\bibfnamefont{S.}~\bibnamefont{Moreno-Flores}},
  \bibinfo{author}{\bibfnamefont{R.}~\bibnamefont{Richert}},
  \bibinfo{author}{\bibfnamefont{H.}~\bibnamefont{Sillescu}}, \bibnamefont{and}
  \bibinfo{author}{\bibfnamefont{D.}~\bibnamefont{Neher}},
  \bibinfo{journal}{Phys. Rev. E} \textbf{\bibinfo{volume}{61}},
  \bibinfo{pages}{1755} (\bibinfo{year}{2000}{\natexlab{a}}).

\bibitem[{\citenamefont{Bauer et~al.}(2000{\natexlab{b}})\citenamefont{Bauer,
  Richert, B\"ohmer, and Christensen}}]{bauer01}
\bibinfo{author}{\bibfnamefont{C.}~\bibnamefont{Bauer}},
  \bibinfo{author}{\bibfnamefont{R.}~\bibnamefont{Richert}},
  \bibinfo{author}{\bibfnamefont{R.}~\bibnamefont{B\"ohmer}}, \bibnamefont{and}
  \bibinfo{author}{\bibfnamefont{T.}~\bibnamefont{Christensen}},
  \bibinfo{journal}{J. Non-Cryst Solids} \textbf{\bibinfo{volume}{262}},
  \bibinfo{pages}{276} (\bibinfo{year}{2000}{\natexlab{b}}).

\bibitem[{\citenamefont{Fukao and Miyamoto}(2001)}]{fukao01}
\bibinfo{author}{\bibfnamefont{K.}~\bibnamefont{Fukao}} \bibnamefont{and}
  \bibinfo{author}{\bibfnamefont{Y.}~\bibnamefont{Miyamoto}},
  \bibinfo{journal}{Phys. Rev. E} \textbf{\bibinfo{volume}{64}},
  \bibinfo{pages}{011803} (\bibinfo{year}{2001}).

\bibitem[{\citenamefont{Meingast et~al.}(1996)\citenamefont{Meingast, Haluska,
  and Kuzmany}}]{meingast96}
\bibinfo{author}{\bibfnamefont{C.}~\bibnamefont{Meingast}},
  \bibinfo{author}{\bibfnamefont{M.}~\bibnamefont{Haluska}}, \bibnamefont{and}
  \bibinfo{author}{\bibfnamefont{H.}~\bibnamefont{Kuzmany}},
  \bibinfo{journal}{J. Non-Cryst. Solids} \textbf{\bibinfo{volume}{201}},
  \bibinfo{pages}{167} (\bibinfo{year}{1996}).

\bibitem[{\citenamefont{Fukao and Miyamoto}(1999)}]{fukao99}
\bibinfo{author}{\bibfnamefont{K.}~\bibnamefont{Fukao}} \bibnamefont{and}
  \bibinfo{author}{\bibfnamefont{Y.}~\bibnamefont{Miyamoto}},
  \bibinfo{journal}{Europhys. Lett.} \textbf{\bibinfo{volume}{46}},
  \bibinfo{pages}{649} (\bibinfo{year}{1999}).

\bibitem[{\citenamefont{Serghei et~al.}(2006)\citenamefont{Serghei, Mikhailova,
  Eichhorn, Voit, and Kremer}}]{serghei06}
\bibinfo{author}{\bibfnamefont{A.}~\bibnamefont{Serghei}},
  \bibinfo{author}{\bibfnamefont{Y.}~\bibnamefont{Mikhailova}},
  \bibinfo{author}{\bibfnamefont{K.~J.} \bibnamefont{Eichhorn}},
  \bibinfo{author}{\bibfnamefont{B.}~\bibnamefont{Voit}}, \bibnamefont{and}
  \bibinfo{author}{\bibfnamefont{F.}~\bibnamefont{Kremer}},
  \bibinfo{journal}{J. Polymer Sci. B-Polymer Phys.}
  \textbf{\bibinfo{volume}{44}}, \bibinfo{pages}{3006} (\bibinfo{year}{2006}).

\bibitem[{\citenamefont{Oh and Green}(2009)}]{oh09}
\bibinfo{author}{\bibfnamefont{H.}~\bibnamefont{Oh}} \bibnamefont{and}
  \bibinfo{author}{\bibfnamefont{P.~F.} \bibnamefont{Green}},
  \bibinfo{journal}{Nature Materials} \textbf{\bibinfo{volume}{8}},
  \bibinfo{pages}{139} (\bibinfo{year}{2009}).

\bibitem[{\citenamefont{B\"ottcher}(1973)}]{bottcher}
\bibinfo{author}{\bibfnamefont{C.~J.~F.} \bibnamefont{B\"ottcher}},
  \emph{\bibinfo{title}{Theory of electric polarization}},
  vol.~\bibinfo{volume}{1} (\bibinfo{publisher}{Elsevier Scientific Publishing
  Company}, \bibinfo{year}{1973}), \bibinfo{edition}{2nd} ed.

\bibitem[{\citenamefont{Niss et~al.}(2005)\citenamefont{Niss, Jakobsen, and
  Olsen}}]{niss05}
\bibinfo{author}{\bibfnamefont{K.}~\bibnamefont{Niss}},
  \bibinfo{author}{\bibfnamefont{B.}~\bibnamefont{Jakobsen}}, \bibnamefont{and}
  \bibinfo{author}{\bibfnamefont{N.~B.} \bibnamefont{Olsen}},
  \bibinfo{journal}{J. Chem. Phys.} \textbf{\bibinfo{volume}{123}}
  (\bibinfo{year}{2005}).

\bibitem[{\citenamefont{Igarashi
  et~al.}(2008{\natexlab{a}})\citenamefont{Igarashi, Christensen, Larsen,
  Olsen, Pedersen, Rasmussen, and Dyre}}]{igarashi08}
\bibinfo{author}{\bibfnamefont{B.}~\bibnamefont{Igarashi}},
  \bibinfo{author}{\bibfnamefont{T.}~\bibnamefont{Christensen}},
  \bibinfo{author}{\bibfnamefont{E.~H.} \bibnamefont{Larsen}},
  \bibinfo{author}{\bibfnamefont{N.~B.} \bibnamefont{Olsen}},
  \bibinfo{author}{\bibfnamefont{I.~H.} \bibnamefont{Pedersen}},
  \bibinfo{author}{\bibfnamefont{T.}~\bibnamefont{Rasmussen}},
  \bibnamefont{and} \bibinfo{author}{\bibfnamefont{J.~C.} \bibnamefont{Dyre}},
  \bibinfo{journal}{Rev. Sci. Instrum.} \textbf{\bibinfo{volume}{79}},
  \bibinfo{pages}{045105} (\bibinfo{year}{2008}{\natexlab{a}}).

\bibitem[{\citenamefont{Hecksher et~al.}(2010)\citenamefont{Hecksher, Olsen,
  Niss, and Dyre}}]{hecksher10}
\bibinfo{author}{\bibfnamefont{T.}~\bibnamefont{Hecksher}},
  \bibinfo{author}{\bibfnamefont{N.~B.} \bibnamefont{Olsen}},
  \bibinfo{author}{\bibfnamefont{K.}~\bibnamefont{Niss}}, \bibnamefont{and}
  \bibinfo{author}{\bibfnamefont{J.~C.} \bibnamefont{Dyre}},
  \bibinfo{journal}{J. Chem. Phys} \textbf{\bibinfo{volume}{133}},
  \bibinfo{pages}{174514} (\bibinfo{year}{2010}).

\bibitem[{\citenamefont{Davidson et~al.}(1992)\citenamefont{Davidson, Bastian,
  and Markley}}]{davidson92}
\bibinfo{author}{\bibfnamefont{M.}~\bibnamefont{Davidson}},
  \bibinfo{author}{\bibfnamefont{S.}~\bibnamefont{Bastian}}, \bibnamefont{and}
  \bibinfo{author}{\bibfnamefont{F.}~\bibnamefont{Markley}}, in
  \emph{\bibinfo{booktitle}{FERMILAB-Conf-92/100}}
  (\bibinfo{organization}{Fermi National Accelerator Laboratory},
  \bibinfo{year}{1992}).

\bibitem[{\citenamefont{Lautrup}(2005)}]{lautrup}
\bibinfo{author}{\bibfnamefont{B.}~\bibnamefont{Lautrup}},
  \emph{\bibinfo{title}{Physics of Continuous Matter}}
  (\bibinfo{publisher}{IoP, Institute of Physics Publishing},
  \bibinfo{year}{2005}).

\bibitem[{\citenamefont{Wallace et~al.}(1995)\citenamefont{Wallace, van\
  Zanten, and Wu}}]{wallace95}
\bibinfo{author}{\bibfnamefont{W.~E.} \bibnamefont{Wallace}},
  \bibinfo{author}{\bibfnamefont{J.~H.} \bibnamefont{van\ Zanten}},
  \bibnamefont{and} \bibinfo{author}{\bibfnamefont{W.~L.} \bibnamefont{Wu}},
  \bibinfo{journal}{Phys. Rev. E} \textbf{\bibinfo{volume}{52}},
  \bibinfo{pages}{3329} (\bibinfo{year}{1995}).

\bibitem[{\citenamefont{Christensen et~al.}(2007)\citenamefont{Christensen,
  Olsen, and Dyre}}]{christensen07}
\bibinfo{author}{\bibfnamefont{T.}~\bibnamefont{Christensen}},
  \bibinfo{author}{\bibfnamefont{N.~B.} \bibnamefont{Olsen}}, \bibnamefont{and}
  \bibinfo{author}{\bibfnamefont{J.~C.} \bibnamefont{Dyre}},
  \bibinfo{journal}{Phys. Rev. E} \textbf{\bibinfo{volume}{75}},
  \bibinfo{pages}{041502} (\bibinfo{year}{2007}).

\bibitem[{\citenamefont{Gundermann}(2012)}]{ditteExp}
\bibinfo{author}{\bibfnamefont{D.}~\bibnamefont{Gundermann}}, Ph.D. thesis,
  \bibinfo{school}{Roskilde University} (\bibinfo{year}{2012}).

\bibitem[{There is a small step of iteration involved in the data treatment
  here, since we use the expansion coefficient we find to get a more precise
  value of it.()}]{note}
There is a small step of iteration involved in the data treatment here, since
  we use the expansion coefficient we find to get a more precise value of it.

\bibitem[{\citenamefont{Sa\u{g}lanmak et~al.}(2010)\citenamefont{Sa\u{g}lanmak,
  Nielsen, Olsen, Dyre, and Niss}}]{saglanmak}
\bibinfo{author}{\bibfnamefont{N.}~\bibnamefont{Sa\u{g}lanmak}},
  \bibinfo{author}{\bibfnamefont{A.~I.} \bibnamefont{Nielsen}},
  \bibinfo{author}{\bibfnamefont{N.~B.} \bibnamefont{Olsen}},
  \bibinfo{author}{\bibfnamefont{J.~C.} \bibnamefont{Dyre}}, \bibnamefont{and}
  \bibinfo{author}{\bibfnamefont{K.}~\bibnamefont{Niss}}, \bibinfo{journal}{J.
  Chem. Phys} \textbf{\bibinfo{volume}{132}}, \bibinfo{pages}{024503}
  (\bibinfo{year}{2010}).

\bibitem[{\citenamefont{Christensen and Olsen}(1995)}]{christensen95}
\bibinfo{author}{\bibfnamefont{T.}~\bibnamefont{Christensen}} \bibnamefont{and}
  \bibinfo{author}{\bibfnamefont{N.~B.} \bibnamefont{Olsen}},
  \bibinfo{journal}{Rev. Sci. Instr.} \textbf{\bibinfo{volume}{66}},
  \bibinfo{pages}{5019} (\bibinfo{year}{1995}).

\bibitem[{\citenamefont{Christensen and Olsen}(1994)}]{christensen94}
\bibinfo{author}{\bibfnamefont{T.}~\bibnamefont{Christensen}} \bibnamefont{and}
  \bibinfo{author}{\bibfnamefont{N.~B.} \bibnamefont{Olsen}},
  \bibinfo{journal}{Phys. Rev. B} \textbf{\bibinfo{volume}{49}},
  \bibinfo{pages}{15396} (\bibinfo{year}{1994}).

\bibitem[{\citenamefont{Jakobsen et~al.}(2010)\citenamefont{Jakobsen, Olsen,
  and Christensen}}]{jakobsen10}
\bibinfo{author}{\bibfnamefont{B.}~\bibnamefont{Jakobsen}},
  \bibinfo{author}{\bibfnamefont{N.~B.} \bibnamefont{Olsen}}, \bibnamefont{and}
  \bibinfo{author}{\bibfnamefont{T.}~\bibnamefont{Christensen}},
  \bibinfo{journal}{Phys. Rev. E tror jeg (accepted)}
  \textbf{\bibinfo{volume}{??}} (\bibinfo{year}{2010}).

\bibitem[{\citenamefont{Igarashi
  et~al.}(2008{\natexlab{b}})\citenamefont{Igarashi, Christensen, Larsen,
  Olsen, Pedersen, Rasmussen, and Dyre}}]{igarashi08b}
\bibinfo{author}{\bibfnamefont{B.}~\bibnamefont{Igarashi}},
  \bibinfo{author}{\bibfnamefont{T.}~\bibnamefont{Christensen}},
  \bibinfo{author}{\bibfnamefont{E.~H.} \bibnamefont{Larsen}},
  \bibinfo{author}{\bibfnamefont{N.~B.} \bibnamefont{Olsen}},
  \bibinfo{author}{\bibfnamefont{I.~H.} \bibnamefont{Pedersen}},
  \bibinfo{author}{\bibfnamefont{T.}~\bibnamefont{Rasmussen}},
  \bibnamefont{and} \bibinfo{author}{\bibfnamefont{J.~C.} \bibnamefont{Dyre}},
  \bibinfo{journal}{Rev. Sci. Instrum.} \textbf{\bibinfo{volume}{79}},
  \bibinfo{pages}{045106} (\bibinfo{year}{2008}{\natexlab{b}}).

\bibitem[{\citenamefont{Jakobsen et~al.}(2011)\citenamefont{Jakobsen, Hecksher,
  Niss, Christensen, Olsen, and Dyre}}]{jakobsen11}
\bibinfo{author}{\bibfnamefont{B.}~\bibnamefont{Jakobsen}},
  \bibinfo{author}{\bibfnamefont{T.}~\bibnamefont{Hecksher}},
  \bibinfo{author}{\bibfnamefont{K.}~\bibnamefont{Niss}},
  \bibinfo{author}{\bibfnamefont{T.}~\bibnamefont{Christensen}},
  \bibinfo{author}{\bibfnamefont{N.~B.} \bibnamefont{Olsen}}, \bibnamefont{and}
  \bibinfo{author}{\bibfnamefont{J.~C.} \bibnamefont{Dyre}},
  \bibinfo{journal}{arXiv:1106.0227v1 [cond-mat.soft]}  (\bibinfo{year}{2011}).

\bibitem[{\citenamefont{Gundermann et~al.}(2011)\citenamefont{Gundermann,
  Pedersen, Hecksher, Bailey, Jakobsen, Christensen, Olsen, Schrøder,
  Fragiadakis, Casalini et~al.}}]{gundermann11}
\bibinfo{author}{\bibfnamefont{D.}~\bibnamefont{Gundermann}},
  \bibinfo{author}{\bibfnamefont{U.~R.} \bibnamefont{Pedersen}},
  \bibinfo{author}{\bibfnamefont{T.}~\bibnamefont{Hecksher}},
  \bibinfo{author}{\bibfnamefont{N.~P.} \bibnamefont{Bailey}},
  \bibinfo{author}{\bibfnamefont{B.}~\bibnamefont{Jakobsen}},
  \bibinfo{author}{\bibfnamefont{T.}~\bibnamefont{Christensen}},
  \bibinfo{author}{\bibfnamefont{N.~B.} \bibnamefont{Olsen}},
  \bibinfo{author}{\bibfnamefont{T.~B.} \bibnamefont{Schrøder}},
  \bibinfo{author}{\bibfnamefont{D.}~\bibnamefont{Fragiadakis}},
  \bibinfo{author}{\bibfnamefont{R.}~\bibnamefont{Casalini}},
  \bibnamefont{et~al.}, \bibinfo{journal}{Nature Physics}
  \textbf{\bibinfo{volume}{7}} (\bibinfo{year}{2011}).

\bibitem[{\citenamefont{Doi and Edwards}(1986)}]{doi}
\bibinfo{author}{\bibfnamefont{M.}~\bibnamefont{Doi}} \bibnamefont{and}
  \bibinfo{author}{\bibfnamefont{S.~F.} \bibnamefont{Edwards}},
  \emph{\bibinfo{title}{The Theory of Polymer Dynamics}}
  (\bibinfo{publisher}{Oxford University Press}, \bibinfo{year}{1986}).

\bibitem[{\citenamefont{Nielsen and Dyre}(1996)}]{nielsen96}
\bibinfo{author}{\bibfnamefont{J.~K.} \bibnamefont{Nielsen}} \bibnamefont{and}
  \bibinfo{author}{\bibfnamefont{J.~C.} \bibnamefont{Dyre}},
  \bibinfo{journal}{Phys. Rev. B} \textbf{\bibinfo{volume}{54}},
  \bibinfo{pages}{15754} (\bibinfo{year}{1996}).

\end{thebibliography}
\end{document}